\documentclass[useAMS,usenatbib]{mn2e}

\usepackage{times,psfig,epsfig} 
\usepackage{rotating, graphicx,dsfont}
\usepackage{color}
\usepackage{amssymb, amsmath}

\usepackage[utf8]{inputenc}
\usepackage{lipsum}
\usepackage{pifont}
\usepackage{amsmath, amssymb, bm}
\usepackage{balance}
\usepackage{multirow}
\usepackage{enumitem}

\newcommand{\be}{\begin{equation}}
\newcommand{\ee}{\end{equation}}
\newcommand{\ba}{\begin{eqnarray}}
\newcommand{\ea}{\end{eqnarray}}

\title{Neutral hydrogen in galaxy clusters: impact of AGN feedback and implications for intensity mapping}

\author[F. Villaescusa-Navarro et al.]
{Francisco Villaescusa-Navarro$^{1,2}$\thanks{e-mail: villaescusa@oats.inaf.it}, 
Susana Planelles$^{3,4}$, 
Stefano Borgani$^{1,2,4}$, 
Matteo Viel$^{1,2}$, 
\newauthor
Elena Rasia$^{1,5}$, 
Giuseppe Murante$^{1}$,
Klaus Dolag$^{6,7}$,
Lisa K. Steinborn$^{6}$,
Veronica Biffi$^{1,4}$,
\newauthor
Alexander M. Beck$^{6}$,
Cinthia Ragone-Figueroa$^{8}$
 \\~\\
\footnotesize 
$^1$ INAF, Osservatorio Astronomico di Trieste, via Tiepolo 11, I-34131 Trieste, Italy\\ 
$^2$ INFN -- National Institute for Nuclear Physics, Via Valerio 2, I-34127 Trieste, Italy\\ 
$^3$ Departament d'Astronomia i Astrofisica, Universitat de Valencia, c/ Dr. Moliner 50, E-46100 - Burjassot, Valencia, Spain\\
$^4$ Astronomy Unit, Department of Physics, University of Trieste, via Tiepolo 11, I-34131 Trieste, Italy\\
$^5$ Department of Physics, University of Michigan, 450 Church St., Ann Arbor, MI 48109, USA\\
$^6$ Universit$\ddot{a}$ts-Sternwarte M$\ddot{u}$nchen, Scheinerstr.1, D-81679 M$\ddot{u}$nchen, Germany\\
$^7$ Max-Plank-Institut f$\ddot{u}$r Astrophysik, Karl-Schwarzschild Strasse 1, D-85740 Garching, Germany\\
$^8$ Instituto de Astronomia Teorica y Experimental (IATE), Consejo Nacional de Investigaciones Cientificas y
Tecnicas de la Republica Argentina (CONICET),\\ ~~Observatorio Astronomico, Universidad Nacional de
Cordoba, Laprida 854, X5000BGR, Cordoba, Argentina
}

\begin{document}
\maketitle 

\begin{abstract}
By means of zoom-in hydrodynamic simulations we quantify the amount of neutral hydrogen (HI) hosted by groups and clusters of galaxies. Our simulations, which are based on an improved formulation of smoothed particle hydrodynamics (SPH), include radiative cooling, star formation, metal enrichment and supernova feedback, and can be split in two different groups, depending on whether feedback from active galactic nuclei (AGN) is turned on or off. Simulations are analyzed to account for HI self-shielding and the presence of molecular hydrogen. We find that the mass in neutral hydrogen of dark matter halos monotonically increases with the halo mass and can be well described by a power-law of the form $M_{\rm HI}(M,z)\propto M^{3/4}$. Our results point out that AGN feedback reduces both the total halo mass and its HI mass, although it is more efficient in removing HI. We conclude that AGN feedback reduces the neutral hydrogen mass of a given halo by $\sim50\%$, with a weak dependence on halo mass and redshift. The spatial distribution of neutral hydrogen within halos is also affected by AGN feedback, whose effect is to decrease the fraction of HI that resides in the halo inner regions. By extrapolating our results to halos not resolved in our simulations we derive astrophysical implications from the measurements of $\Omega_{\rm HI}(z)$: halos with circular velocities larger than $\sim25~{\rm km/s}$ are needed to host HI in order to reproduce observations. We find that only the model with AGN feedback is capable of reproducing the value of $\Omega_{\rm HI}b_{\rm HI}$ derived from available 21cm intensity mapping observations. 
\end{abstract} 
 

\section{Introduction}
\label{sec:introduction}

The formation and evolution of galaxies is an extremely complicated process that we do not fully understand yet. In the standard picture, gas falls within the gravitational potential wells of dark matter halos, where it cools down and eventually forms stars. Galaxies grow mainly by accreting gas from the intergalactic medium (IGM). That gas, which is mainly ionized, becomes neutral once its density is high enough to be self-shielded against the exterior radiation. Neutral hydrogen (HI), is one of the major gas components in galaxies, and it has been shown that strong correlations show up among the HI content and stellar mass \citep{Catinella_2010, Cortese_2011, Huang_2012}. This is surprising because stars form from the collapse and fragmentation of molecular hydrogen (${\rm H}_2$) clouds, which are created, under proper conditions, from HI and, therefore, the presence of neutral hydrogen does not imply that star formation is taking place.

From observations we know that galaxies can be classified in two different groups. One group is dominated by blue star forming galaxies, which are rich in cold gas (late-type) whereas the other group (early-type) contains red passive galaxies which host, on average, low gas fraction. The physical mechanisms responsible for this bimodality are not yet fully understood. Among the different mechanisms that can quench star formation there are mergers \citep{Toomre_1972}, gas stripping by ram-pressure \citep{Gunn_Gott_1972, Yoon_2015} and AGN feedback \citep{Ho_2008, Hughes_2009,Leslie_2015}. The relative importance of the different processes is also not well known. 

In order to improve our understanding of the physical processes responsible of galaxy formation and evolution, it is thus very important to perform observations which not only tell us about the star formation rate (SFR) and the stellar mass, but also which inform us on the mass in neutral and molecular hydrogen, being those the main constituents of the interstellar medium (ISM). One of the purposes of this paper is to investigate the impact of AGN feedback on the HI content in groups and clusters of galaxies.

Besides using HI to improve our understanding of galaxy formation and evolution, neutral hydrogen plays a key role in cosmology, since it can be used as a tracer of the large scale structure of the Universe. Therefore, it is very important to properly model the spatial distribution of neutral hydrogen, in the post-reionization epoch, as it will be sampled, either by HI-selected galaxies \citep{Yahya_2015,2015aska.confE..21S} or via intensity mapping \citep{Battye:2004re, Chang_2010, Masui_2013, Switzer_2013, Santos_2015}, by radio-telescopes as the Giant Meterwave Radio Telescope (GMRT)\footnote{http://gmrt.ncra.tifr.res.in/}, the Ooty Radio Telescope (ORT)\footnote{http://rac.ncra.tifr.res.in/}, the Low-Frequency Array (LOFAR)\footnote{http://www.lofar.org/}, the Murchison Wide-field Array (MWA)\footnote{http://www.mwatelescope.org/}, the Canadian Hydrogen Intensity Mapping Experiment (CHIME)\footnote{http://chime.phas.ubc.ca/}, the Five hundred meter Aperture Spherical Telescope (FAST)\footnote{http://fast.bao.ac.cn/en/}, ASKAP (The Australian Square Kilometer Array Pathfinder)\footnote{http://www.atnf.csiro.au/projects/askap/index.html}, MeerKAT (The South African Square Kilometer Array Pathfinder)\footnote{http://www.ska.ac.za/meerkat/} and the future SKA (The Square Kilometer Array)\footnote{https://www.skatelescope.org/}. 

Intensity mapping is a new technique to sample the large scale structure of the Universe which consists in performing a low angular resolution survey where the 21cm emission from individual galaxies is not resolved. By using this technique, the radio telescopes will just measure the integrated emission from neutral hydrogen from many unresolved galaxies. The underlying idea is that on large scales, fluctuations in the integrated 21cm signal will follow that of the underlying matter \citep{Bharadwaj_2001A, Bharadwaj_2001B,Battye:2004re,McQuinn_2006,Chang_2008,Loeb_Wyithe_2008, Bull_2015, Santos_2015, Villaescusa-Navarro_2015a}. Thus, intensity mapping represents a different way to sample the large scale structure of the Universe and it is expected to revolutionize cosmology given the very large volumes it can be sampled and the spectroscopic nature of the measurements \citep{Bull_2015,Santos_2015,Villaescusa-Navarro_2015a,Alonso_2015a}.

The function $M_{\rm HI}(M,z)$, which represents the average neutral hydrogen mass hosted by a dark matter halo of mass $M$ at redshift $z$, plays a key role in 21cm cosmology, since the shape and amplitude of the 21cm power spectrum, on large, linear scales, is completely determined by that function. We now briefly explain  the reason of this. 

On large scales, the amplitude of the 21cm power spectrum can be written as (see Villaescusa-Navarro et al. 2015 for a detailed discussion)
\begin{eqnarray}
P_{\rm 21cm}(k,z)&=&\overline{\delta T_b}^2(z)b_{\rm HI}^2(z)\left(1+\frac{2}{3}\beta(z)+\frac{1}{5}\beta^2(z)\right)\\ \nonumber
&&\times ~P_{\rm m}(k,z)~, 
\label{eq:P_21cm}
\end{eqnarray}
where $\beta(z)=f(z)/b_{\rm HI}(z)$ is the redshift-space distortion parameter and the third factor on the right-hand side of the above equation arises from the Kaiser formula \citep{Kaiser_1987}. $b_{\rm HI}(z)$ represents the bias of the neutral hydrogen, $P_{\rm m}(k,z)$ denotes the linear matter power spectrum and the mean brightness temperature, $\overline{\delta T_b}(z)$, is given by
\begin{equation}
\overline{\delta T_b}(z) = 189\left(\frac{H_0(1+z)^2}{H(z)}\right)\Omega_{\rm HI}(z)h~{\rm mK}~,
\label{eq:delta_Tb}
\end{equation}
where $H(z)$ and $H_0$ are the value of the Hubble parameter at redshifts $z$ and $0$, respectively. $h$ represents the value of $H_0$ in units of $100~{\rm km~s^{-1} Mpc^{-1}}$ and $\Omega_{\rm HI}(z)$ is the ratio between the comoving neutral hydrogen density at redshift $z$ to the Universe critical density at $z=0$, $\rho^0_{\rm c}$. 

Therefore, for a given cosmological model, the amplitude of the 21cm signal depends on the amount of neutral hydrogen, via $\Omega_{\rm HI}(z)$, but also on the way the HI is distributed among different halos, i.e. on the neutral hydrogen bias, $b_{\rm HI}(z)$. Contrary to what observers usually measure (the HI mass within galaxies), the important quantity for cosmology is the function $M_{\rm HI}(M,z)$, which gives the average HI mass hosted by a halo of mass $M$ at redshift $z$. Given that function, it is straightforward to compute both $\Omega_{\rm HI}(z)$ and $b_{\rm HI}(z)$ as
\begin{eqnarray}
\Omega_{\rm HI}(z) & = & \frac{1}{\rho^0_{\rm c}}\int_0^\infty n(M,z)M_{\rm HI}(M,z)dM~,\\~
\nonumber\\
b_{\rm HI}(z) & = & \frac{\int_0^\infty n(M,z)b(M,z)M_{\rm HI}(M,z)dM}{\int_0^\infty n(M,z)M_{\rm HI}(M,z)dM}~.
\label{eq:bias_HI}
\end{eqnarray}
where $n(M,z)$ and $b(M,z)$ are the halo mass function and halo bias, respectively. In other words, for a given cosmological model, the function $M_{\rm HI}(M,z)$ completely determines the shape and amplitude of the 21cm power spectrum on large scales.

The HI mass residing in galaxy clusters plays a fundamental role in determining the value of the HI bias  \cite[see][for an observational study on the cold atomic gas content of galaxies in groups and clusters]{Yoon_2015}. The reason is that clusters of galaxies are very biased objects, and therefore, if a significant amount of neutral hydrogen is found within them, they can substantially enhance the value of $b_{\rm HI}(z)$. This can be rephrased saying that the high mass end of the function $M_{\rm HI}(M,z)$ is very important since the HI bias strongly depends on it.

At $z\simeq0$, the bias of HI selected galaxies\footnote{Notice that this bias is different to the one defined in Eq. \ref{eq:bias_HI}. The bias of HI selected galaxies is defined as the value of $b_{\rm HI,gal}^2(r)=\xi_{\rm HI,gal}(r)/\xi_{\rm m}(r)$ on large scales, where $\xi_{\rm HI,gal}(r)$ and $\xi_{\rm m}$ represent the 2-point correlation function of the HI selected galaxies and the underlying matter, respectively.} have been measured from the HI Parkes All-Sky Survey \citep[HIPASS,][]{Barnes_2001} obtaining a value of $b_{\rm HI,gal}=0.7\pm0.1$, in good agreement with the one measured from the Arecibo Legacy Fast ALFA Survey
\citep[ALFALFA,][]{Giovanelli_2005,Martin_2012} \citep[see however][for analytical estimates of the HI bias at $z=0$ and its redshift evolution]{Marin_2010}. At $\langle z \rangle=0.8$, 21cm intensity mapping observations with the Green Bank Telescope (GBT) have been used to measure the product $b_{\rm HI}(z)\times\Omega_{\rm HI}(z)$, finding a value of $6.2^{+2.3}_{-1.5}\times10^{-4}$ \citep{Chang_2010,Masui_2013,Switzer_2013}. Finally, at $\langle z \rangle=2.3$ the bias of damped Lyman-$\alpha$ absorbers (DLAs) has been recently estimated in \citet{Font_2012} to be $b_{\rm DLAs}=(2.17\pm0.20)\beta_F^{0.22}$, where $\beta_F$ is the Ly-$\alpha$ forest redshift distortion parameter whose value is of order 1.

Thus, it is very important to model as best as possible the function $M_{\rm HI}(M,z)$ since the signal-to-noise ratio of the 21cm signal depends on it. On the other hand, the function $M_{\rm HI}(M,z)$ contains important astrophysical information, giving us information on how HI is distributed among the different halos and it can be constrained by combining observations from the Ly$\alpha$-forest, 21cm intensity mapping and so on. 

Theoretical models like the one presented in \cite{Bagla_2010} propose a phenomenological function for $M_{\rm HI}(M,z)$ making the hypothesis that galaxy clusters do not host a significant amount of neutral hydrogen. This hypothesis relies on the fact that observations point out that galaxies in clusters are HI deficient \citep{Solanes_2001, Gavazzi_2005, Gavazzi_2006, Rhys_2012a, Rhys_2012b, Catinella_2013, Denes_2014}. In \cite{Villaescusa-Navarro_2014a} it was shown that this simple model is capable of reproducing extremely well the abundance of DLAs at redshifts $z=[2.4-4]$, while in \cite{Padmanabhan_2015} authors claim that the model successfully reproduces the bias of the HI selected galaxies at $z\simeq0$. In this paper we check the validity of the assumption behind the \citet{Bagla_2010} model.

The purpose of this paper is to investigate, using state-of-the-art zoom-in hydrodynamic simulations, the amount of neutral hydrogen hosted by groups and clusters of galaxies (i.e. the high-mass end of the $M_{\rm HI}(M,z)$ function), its spatial distribution within halos and its evolution with time. We also test the validity of the \citet{Bagla_2010} assumptions by comparing the results of our simulations against the prediction of that model. Moreover, we study the impact of active galactic nuclei feedback on the neutral hydrogen content of clusters and groups and investigate the implications for 21cm intensity mapping.

This paper is organized as follows. In Sec. \ref{sec:simulations} we describe the hydrodynamic simulations used for this work, together with the method used to model the spatial distribution of neutral hydrogen. We investigate in Sec. \ref{sec:results} the mass in neutral hydrogen hosted by groups and clusters, its spatial distribution within halos, its redshift evolution and the impact of AGN feedback. In Sec. \ref{sec:Bagla} we compare our findings against the \citet{Bagla_2010} model and study the consequences for 21cm intensity mapping. We draw the main conclusions and discuss the results of this paper in Sec. \ref{sec:conclusions}. In the Appendix \ref{sec:HIMF} we discuss the distribution of neutral hydrogen among galaxies belonging to groups and clusters and the possible level of numerical contamination.


\section{Simulations}
\label{sec:simulations}

In this section we describe the set of zoom-in hydrodynamic simulations we have used for this work. We then depict the method we employ to model the spatial distribution of neutral hydrogen and the procedure utilized to identify dark matter halos and galaxies.

\subsection{Hydrodynamic simulations}
\label{subsec:simulations}

Here, we provide a short characterization of the set of simulations used in this paper. While we refer to a  future work (Planelles et. al in preparation) for a more detailed description, a first analysis of these set of simulations can be found in the recently submitted paper by \citet[][]{Rasia_2015}.
The set of hydrodynamic simulations  consists in re-simulations  of 29 Lagrangian regions  centered around the 29 most massive halos formed in a larger N-body cosmological simulation \citep[see][for details on the initial conditions]{Bonafede2011}. 
The simulations, performed with the TreePM--SPH {\footnotesize {\sc GADGET-3}} code \citep{Springel_2005}, assume a flat $\Lambda$CDM cosmology with 
$\Omega_{\rm{m}} = 0.24$, $\Omega_{\rm{b}} = 0.04$, $H_0=$~72~km~s$^{-1}$~Mpc$^{-1}$, $n_{\rm{s}}=0.96$ , $\sigma_8 = 0.8$.

The mass resolution for the DM particles and the initial mass of the gas particles are, respectively, $m_{\rm{DM}} = 8.44\times10^8 \, {h}^{-1} M_{\odot}$ and $m_{\rm{gas}} = 1.56\times10^8\, {h}^{-1} M_{\odot}$.
As for the spatial resolution, gravitational force in the re-simulated regions is computed with a Plummer-equivalent softening length of 
$\epsilon =3.75\, {h}^{-1}$ kpc (in physical units at $z < 2$, while fixed in comoving units at $z>2$).

An improved version of the standard SPH scheme, as discussed in  \citet{Beck_2015}, has been  included. This new hydro scheme  includes a number of elements (such as an artificial conduction term, a time-dependent artificial viscosity and a Wendland $C^4$  interpolation kernel) which largely improves the performance of the traditional SPH scheme. Including this new hydro scheme, different sets of simulations, characterized by the inclusion of different sets of baryonic physical processes, have been performed. 
In this work,  our reference simulations, labelled as the AGN simulations, include the effects of the AGN feedback model recently presented in \citet{Steinborn_2015}. In this model, accretion onto super massive black holes (SMBHs) takes place according to the Bondi formula and is Eddington-limited, but in contrast to the original implementation by \citet{Springel_2005_AGN}, we compute the accretion rate separately for cold and hot gas using different boost factors (typically, 100 for cold gas and 10 for hot gas accretion). In the particular set of simulations presented in this paper, as in \citet[][]{Rasia_2015}, we neglect any contribution from hot gas accretion. Furthermore, both mechanical outflows and radiation contribute to the thermal energy. Using variable efficiencies for these two components allows a continuous transition between the quasar and the radio mode. For the radiative feedback we fix the coupling factor to the surrounding gas to $\epsilon_f=0.05$. We refer to  \citet{Steinborn_2015} for a more detailed description of this model.

Besides AGN feedback, our reference model also accounts for the effects of a number of additional processes such as  metallicity-dependent radiative cooling, star formation and supernova (SN) feedback and metal enrichment. These processes have been included as described in  \cite{Planelles_2014}. Briefly, radiative cooling and the presence of the UV/X-ray background radiation  are included according to, respectively, \cite{wiersma_etal09} and \cite{haardt_madau01}. 
The sub-grid model for star formation and its associated feedback is implemented according to the prescription by \cite{springel_hernquist03}. Galactic winds with a velocity of $\sim 350~{\rm km/s}$ characterize the kinetic feedback from SNe. Metal enrichment is also included according to the chemical model by \cite{tornatore_etal07}. We refer to this simulation set as AGN run.

We emphasize that our AGN reference model has been shown to agree with a number of cluster observations and, therefore, to provide a realistic population of clusters. In particular, in the recent work by  \citet[][]{Rasia_2015} it has been shown how this set of simulations produce, in a natural way, the coexistence of cool-core and non-cool-core clusters, with entropy and iron abundance profiles in good agreement with observational profiles. A similar good match with observed data is also obtained for  other cluster properties such as  pressure profiles,  gas and baryon mass fractions or X-ray and SZ scaling relations (see Planelles et al. and Truong et al., both in preparation). Our reference simulation, appears also to reproduce the observed scaling between stellar mass of the host galaxy and BH mass, producing as well a good estimate of BCG masses. 
 
For completeness, besides our reference AGN simulations, we will also analyze another set of radiative simulations, labelled as CSF, which include the same physical processes than our reference model but for which we have turned off AGN feedback. We use these simulations to investigate the impact of AGN feedback by comparing the results from this set against the one from the AGN simulations.

\subsection{Neutral hydrogen distribution}
\label{subsec:HI_distribution}

We need to post-process the output of the simulations to account for two critical processes, which are not followed by our simulations, to properly model the distribution of neutral hydrogen: the HI self-shielding and the formation of molecular hydrogen, H$_2$. 

The procedure used to model the spatial distribution of neutral hydrogen is as follows. First of all, the hydrogen mass that is neutral is computed, for all gas particles in the simulation, assuming ionization equilibrium and estimating the photo-ionization rate using the fitting formula of \citet{Rahmati_2013}. Then, we correct the HI mass obtained above by modeling the presence of molecular hydrogen using the model developed in \citet{Krumholz_2008, Krumholz_2009, McKee_2010}, the so-called KMT model, assuming that only star-forming particles host H$_2$.

The HI self-shielding is taken into account by using the fitting formula of \citet{Rahmati_2013}, that we use to compute the fraction of hydrogen that is in neutral state (i.e. both neutral hydrogen, HI, and molecular hydrogen, H$_2$). We now briefly describe the method used here and refer the reader to \citet{Rahmati_2013} for further details. The amount of hydrogen that is neutral is computed, for each gas particle, assuming ionization equilibrium \citep[see for instance Appendix A2 of][]{Rahmati_2013} where the photo-ionization rate, $\Gamma_{\rm Phot}$, affecting a given gas particle is a function of its hydrogen number density, $n_{\rm H}$
\be
\Gamma_{\rm Phot}=\Gamma_{\rm UVB}\left\{(1-f)\left[1+\left(\frac{n_{\rm H}}{n_0}\right)^\beta\right]^{\alpha_1}+f\left[1+\frac{n_{\rm H}}{n_0}\right]^{\alpha_2}\right\}~,
\ee
where $\Gamma_{\rm UVB}$ is the UV background photo-ionization rate and $n_0$, $\alpha_1$, $\alpha_2$, $\beta$ and $f$ are free-parameters of the fitting formula whose values we take from the Table A1 of \citet{Rahmati_2013}. We interpolate to obtain the value of the above parameters for redshifts not covered by their Table A1. We notice that we are neglecting radiation from local sources \citep{Miralda-Escude_2005, Schaye_2006, Rahmati_sources} and local X-ray radiation from the hot intra-cluster medium (ICM) \citep{Kannan_2015} when computing the mass in neutral hydrogen. In Sec. \ref{sec:conclusions} we discuss the possible consequences of this assumption in our results. 

The presence of molecular hydrogen, H$_2$, is modeled using the KMT analytic model. The molecular hydrogen fraction, $f_{\rm H_2}$, defined as $f_{\rm H_2}=M_{\rm H_2}/M_{\rm NH}$, where $M_{\rm H_2}$ is the mass in molecular hydrogen and $M_{\rm NH}=M_{\rm HI}+M_{\rm H_2}$ is the mass in hydrogen which is neutral (computed as explained in the previous paragraph) is estimated using
\be
f_{\rm H_2} = \left\{ 
  \begin{array}{l l}	
  
    1-\frac{0.75s}{1+0.25s} & \quad \text{if $s<2$}
    \\
    0 & \quad \text{if $s\geqslant2$}\\
  \end{array} \right.
\label{eq:f_H2}
\ee
where 
\be
s=\frac{\log(1+0.6\chi+0.01\chi^2)}{0.6\tau_c}
\ee
with
\begin{eqnarray}
\chi&=&0.756(1+3.1Z^{0.365})\\
\tau_c&=&\Sigma\sigma_d/\mu_{\rm H}
\end{eqnarray}
where $Z$ is the metallicity of the gas particle, self-consistently described in our simulations, in units of the solar metallicity \citep{Allende_Prieto_2001}, $\mu_{\rm H}$ is the mean mass per hydrogen nucleus ($\mu_{\rm H}=2.3\times10^{-24}~{\rm g}$), $\sigma_d$ is the dust cross-section (we take $\sigma_d=Z\times10^{-21}~{\rm cm^2}$) and $\Sigma$ is the gas surface density\footnote{We follow \cite{Dave_2013} and compute the surface density of a given gas particle by multiplying its density by its SPH radius.}. We assume that the extent and profile of the metallicity is governed by the particle SPH radius and kernel, respectively. Notice that we assume that only star-forming particles (i.e. particles with physical densities higher than $\sim0.1~{\rm H/cm}^{-3}$) host H$_2$, thus, the HI mass computed above only needs to be corrected in those particles.

\subsection{Identification of dark matter halos and galaxies}
\label{sec:halos_identification}

Dark matter halos are initially found by applying the Friends-of-Friends (FoF) algorithm \citep{FoF} with a value of the linking length parameter $b=0.16$ on top of the simulation snapshots. Next, the center of a given halo is found by searching the particle with the minimum value of the gravitational potential. Then, the halo radius, $R_{200}$, and its mass, $M$, are computed by requiring that the mean density of a sphere centered on the halo center and with that radius is equal to 200 times the critical density of the Universe at that redshift, $\rho_c(z)$, i.e.
\be
M=\frac{4\pi}{3}\rho_c(z)\triangle_c(z)R_{200}^3
\ee
where $\triangle_c(z)=200$. The above procedure is also used to find $\{M_{500},R_{500}\}$ and $\{R_{2500},M_{2500}\}$, which are defined as above but setting $\triangle_c(z)=500$ and $\triangle_c(z)=2500$, respectively. Finally, galaxies are identified by means of the {\sc SUBFIND} algorithm \citep{Subfind,Dolag_2009}. At $z=0$, the catalogues of each, the AGN and CSF, simulations contain approximately 450 groups and clusters.


\section{Neutral hydrogen in groups and clusters: impact of AGN feedback}
\label{sec:results}

\begin{figure}
\centerline{\includegraphics[width=0.5\textwidth]{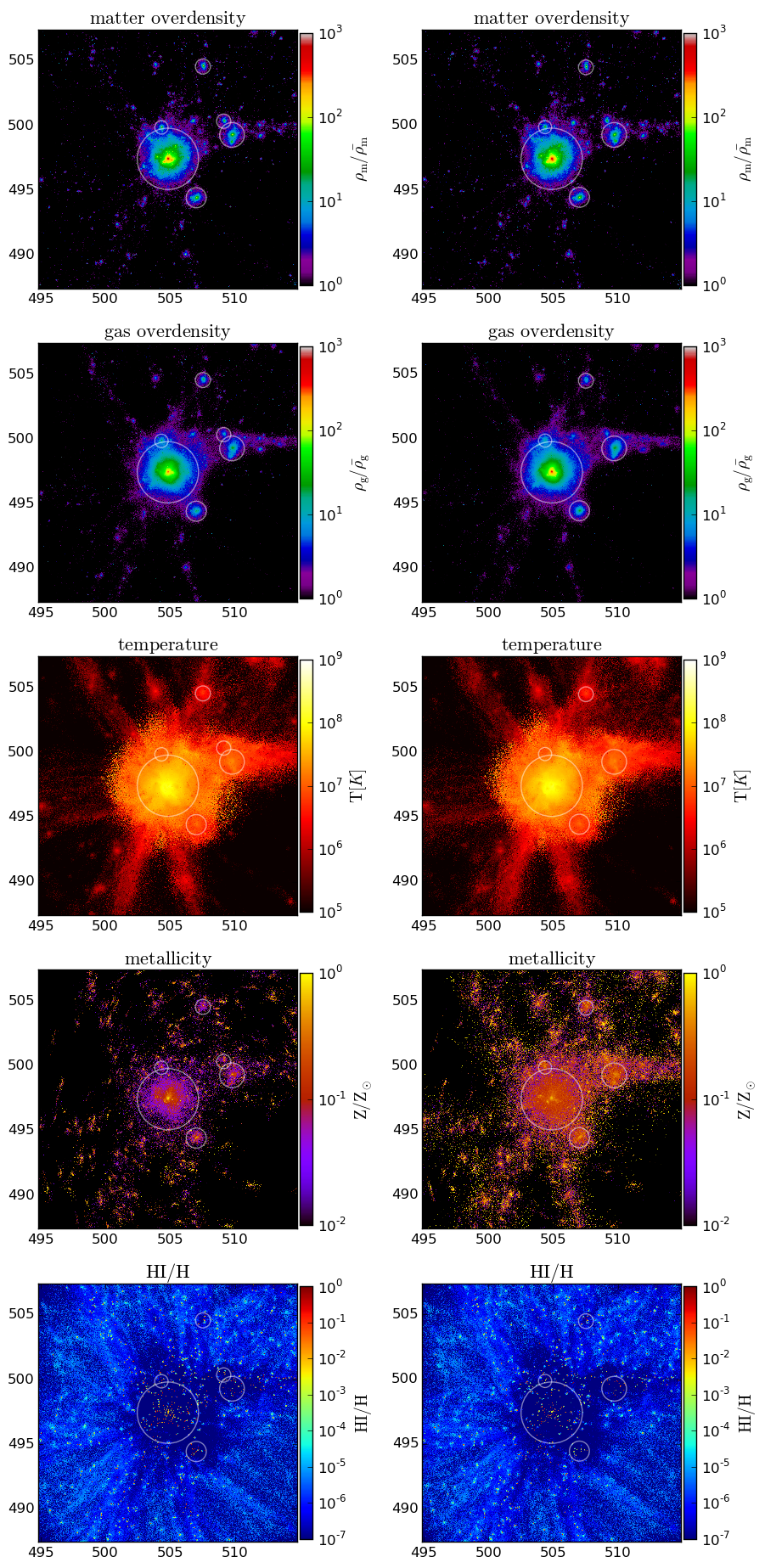}}
\caption{Spatial distribution of matter overdensity (first row), gas overdensity (second row), temperature (third row), metallicity (fourth row) and neutral hydrogen fraction (fifth row) from a slice of 10 $h^{-1}$Mpc width centered in the most massive halo of one of our simulated regions at $z=0$. White circles represent the location and radii of the halos found in that slice. Left and right panels show the results for the simulations CSF and AGN, respectively. The units of the X and Y axes are $h^{-1}$Mpc.}
\label{fig:image}
\end{figure}

In this section we present the main results that we obtain by analyzing the hydrodynamic simulations. In order to have a preliminary visual impression of the results, we show in Fig. \ref{fig:image} the spatial distribution of matter, gas, temperature, metallicity and neutral hydrogen fraction around a massive cluster at $z=0$ in one of our resimulated regions. The left and right columns display the results for the simulations without and with AGN feedback, respectively. 

In subsection \ref{subsec:HI_clusters} we investigate the mass and spatial distribution of HI within groups and clusters of galaxies. We investigate the impact of AGN feedback on the amount and distribution of HI in subsection \ref{subsec:AGN_feedback}. In subsection \ref{subsec:redshift_evolution} we study the evolution with time of our results. Finally, we refer the reader to the Appendix \ref{sec:HIMF} for the results on the neutral hydrogen content in galaxies belonging to groups and clusters of galaxies. In that Appendix we also quantify the level of contamination of our results from spurious numerical artifacts in our simulations.

\subsection{Neutral hydrogen in groups and clusters at $z=0$}
\label{subsec:HI_clusters}

\begin{figure}
\centerline{\includegraphics[width=0.5\textwidth]{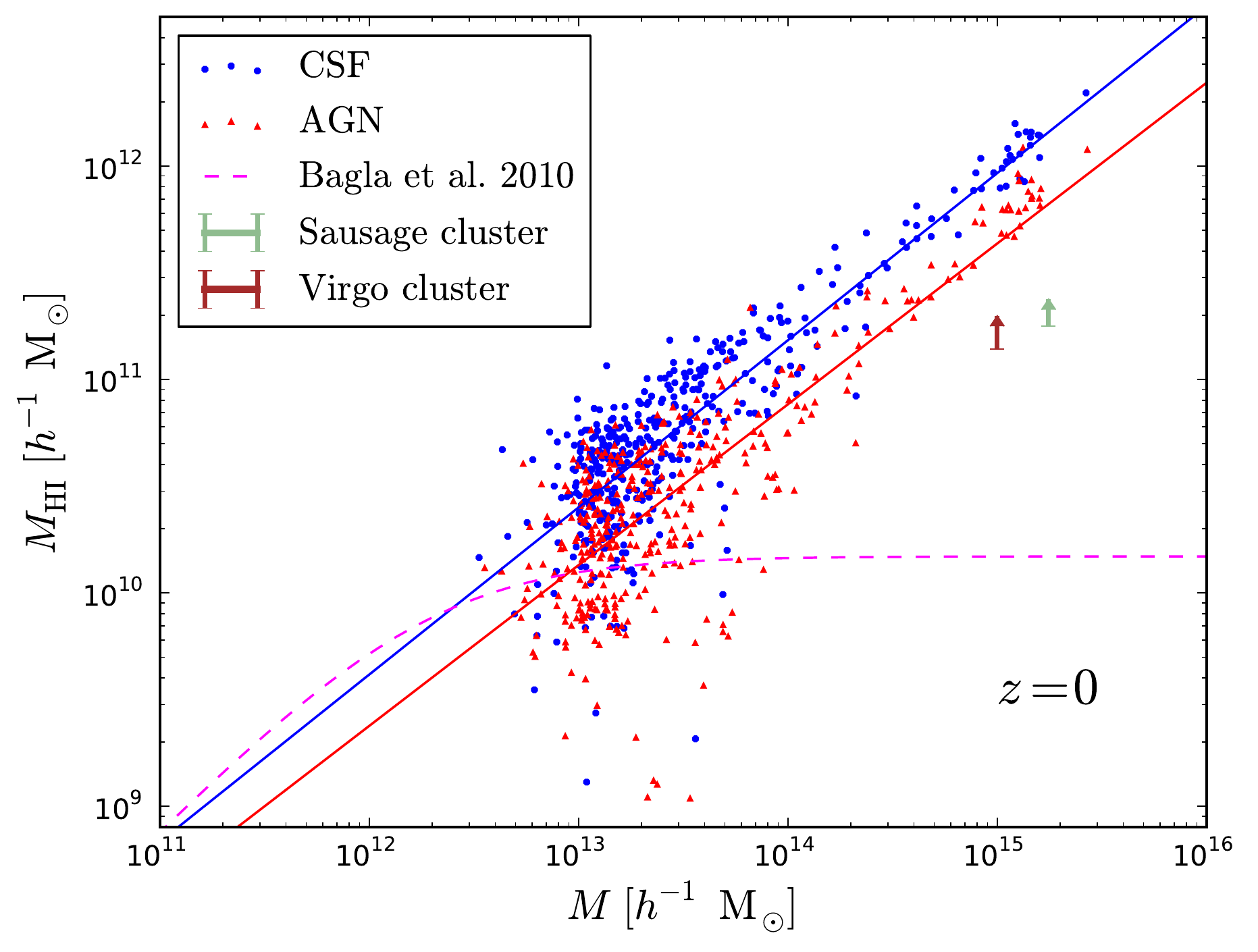}}
\caption{Neutral hydrogen mass within $R_{200}$ as a function of the halo mass, $M$, for the halos of the simulations CSF (blue dots) and AGN (red triangles) at $z=0$. The solid lines show a fit to the results using the function $M_{\rm HI}(M,z)=e^\gamma M^\alpha$. The dashed magenta line displays the prediction of the \citet{Bagla_2010} model. The green and brown arrows represent a lower limit on the neutral hydrogen mass hosted by the Sausage and Virgo clusters, respectively (see text for details).}
\label{fig:M_HI}
\end{figure}

For each halo of the simulations CSF and AGN we have computed the total neutral hydrogen mass within $R_{200}$ and in Fig. \ref{fig:M_HI} we show the HI mass versus the total halo mass, $M_{200}$, at $z=0$. 

\begin{table}
\centering{
{\renewcommand{\arraystretch}{1.4}
\resizebox{8cm}{!}{
\begin{tabular}{|c|c|cc|}
\hline
Simulation & $z$ & $\alpha$ & $\gamma$ \\
\hline
\multirow{6}{*}{CSF} & 0      & $0.78\pm0.02$ & $0.7\pm0.7$\\
                                 & 0.25 & $0.76\pm0.02$ & $1.5\pm0.6$ \\
                                 & 0.5   & $0.75\pm0.02$ & $2.1\pm0.6$ \\	
                                 & 0.8   & $0.73\pm0.02$ & $2.5\pm0.6$ \\	                                 	
                                 & 1      & $0.71\pm0.03$ & $3.2\pm0.8$ \\	
                                 & 1.5   & $0.69\pm0.03$ & $4.1\pm0.8$ \\	
                                 & 2      & $0.59\pm0.03$ & $7.3\pm1.0$ \\	
\hline
\multirow{6}{*}{AGN} & 0      & $0.75\pm0.02$ & $0.9\pm0.8$\\
                                  & 0.25 & $0.77\pm0.02$ & $0.4\pm0.8$ \\
                                  & 0.5   & $0.79\pm0.02$ & $0.1\pm0.8$ \\	
                                  & 0.8   & $0.75\pm0.02$ & $1.3\pm0.6$ \\	                                 	
                                  & 1      & $0.73\pm0.03$ & $2.0\pm0.9$ \\	
                                  & 1.5   & $0.73\pm0.03$ & $2.1\pm1.0$ \\	
                                  & 2      & $0.65\pm0.05$ & $4.7\pm1.5$ \\	
                                  & 0~-~1.5 & $0.75\pm0.01$ & $1.1\pm0.4$ \\	
                                  & 0~-~2 & $0.75\pm0.01$ & $1.3\pm0.4$ \\	
\hline
\end{tabular}}}}
\caption{Best-fit parameters for $M_{\rm HI}(M,z)=e^\gamma M^\alpha$ for the two different simulations and for different redshifts. The quoted values of $\alpha$ and $\gamma$ hold when the masses of both $M_{\rm HI}$ and $M$ are in units of $h^{-1}M_\odot$.} \vspace{-1.5em}
\label{tbl:M_HI}
\end{table}

We find that, on average, the bigger the halo the larger the HI mass it hosts. This behavior takes place independently of whether AGN feedback is switched on or off in the simulations. We find that a simple power law of the form\footnote{Notice that in order to give the correct units to all the terms, this relation should be understood as: $M_{\rm HI}(M,z)/(h^{-1}M_\odot)=e^\gamma (M/(h^{-1}M_\odot))^\alpha$.} $M_{\rm HI}(M,z)=e^\gamma M^\alpha$ can reproduce the mean of our results very well. In Table \ref{tbl:M_HI} we show the best-fit values of the parameters $\alpha$ and $\gamma$ for the simulations CSF and AGN at different redshifts. The solid lines in Fig. \ref{fig:M_HI} display those fits at $z=0$. The physical interpretation of the parameters $\alpha$ and $\gamma$ is straightforward. $\gamma$ represents an overall normalization: the HI mass of a halo of mass $1~h^{-1}M_\odot$ is given by $e^\gamma~h^{-1}M_\odot$. On the other hand $\alpha$ characterizes the slope of the $M_{\rm HI}-M$ relation. 

The green and brown arrows in Fig. \ref{fig:M_HI} represent a lower limit, although close to the actual value, on the overall HI mass of the Sausage and Virgo clusters estimated from \citet{Stroe_2015} and \citet{Gavazzi_2005}, respectively\footnote{For the Sausage cluster we consider that the approximately 100 H$\alpha$ star-forming galaxies detected \citep{Stroe_2014} host an average HI mass equal to $2.5\times10^9~M_\odot$ \citep{Stroe_2015} (Andra Stroe, private communication). We estimate the HI mass of the Virgo cluster by summing the HI masses of the 296 detected galaxies from the table A2 of \citet{Gavazzi_2005}.}. It can be seen that these lower limits are a factor of $\sim3-4$ lower than the typical HI masses we find in the halos of the AGN simulation with similar masses. We remark that the HI masses we measure in simulated galaxy clusters may be partially affected\footnote{Notice that we estimate the numerical contamination for very massive galaxy clusters to be $\sim30\%$. Thus, it is unlikely that differences among our simulations and observations are only due to spurious HI blobs.} by numerical artifacts that can spuriously increase the predicted mass in neutral hydrogen by a factor of 2 (see Appendix \ref{sec:HIMF}). Besides, in our calculations we neglect the radiation from local sources and from the hot ICM. 

One may wonder whether the HI masses we measure for groups of galaxies are biased since we are selecting those halos from regions which contain very massive halos. In order to answer this question we have selected different regions in which the mass of the most massive halo is different and computed the HI mass within groups. We find that the mass in neutral hydrogen in groups does not depend on the mass of the most massive halo in the region. We thus conclude that our results are not biased by limiting our study to the highly biased regions we simulate in this work.

\begin{figure}
\centerline{\includegraphics[width=0.5\textwidth]{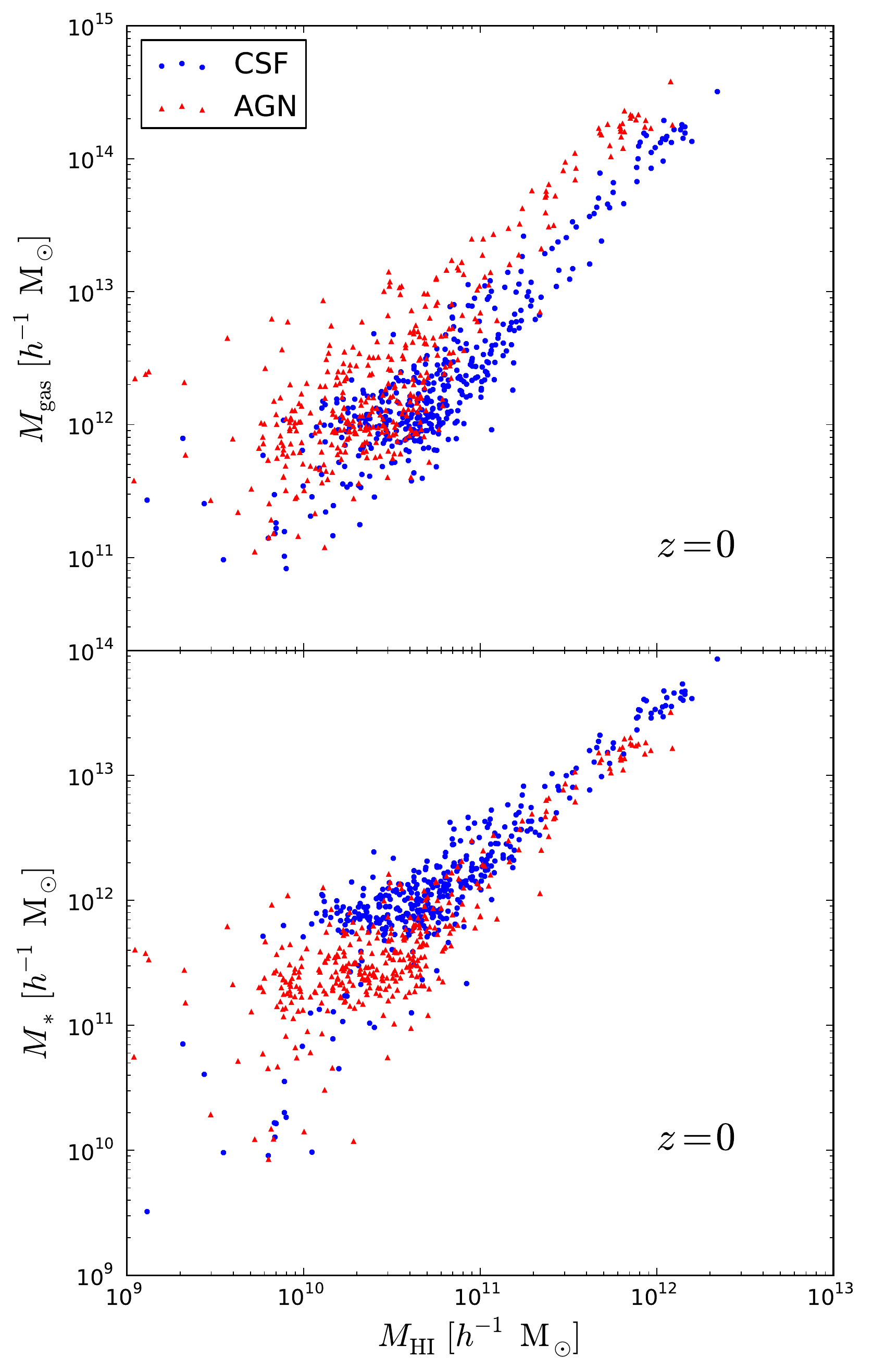}}
\caption{For each halo of the AGN (red) and CSF (blue) simulations at $z=0$ we have computed the gas, stars and neutral hydrogen mass within $R_{200}$. The upper panel shows the gas mass versus the HI mass while the bottom panel displays the stellar mass versus the HI mass.}
\label{fig:HI_correlations}
\end{figure}

We have also studied the correlations between the overall gas, stars and HI mass in halos. For each halo of the CSF and AGN simulations we have computed the HI, gas and stellar mass within its $R_{200}$ at $z=0$. In Fig. \ref{fig:HI_correlations} we show the results. We find that the amount of gas and the stellar mass within halos increases with the HI mass, which, on the other hand increases with the mass of the halo. This just reflects the fact that the larger the halo the more CDM, stars, gas and neutral hydrogen it contains.

\begin{figure*}
\centerline{\includegraphics[width=0.8\textwidth]{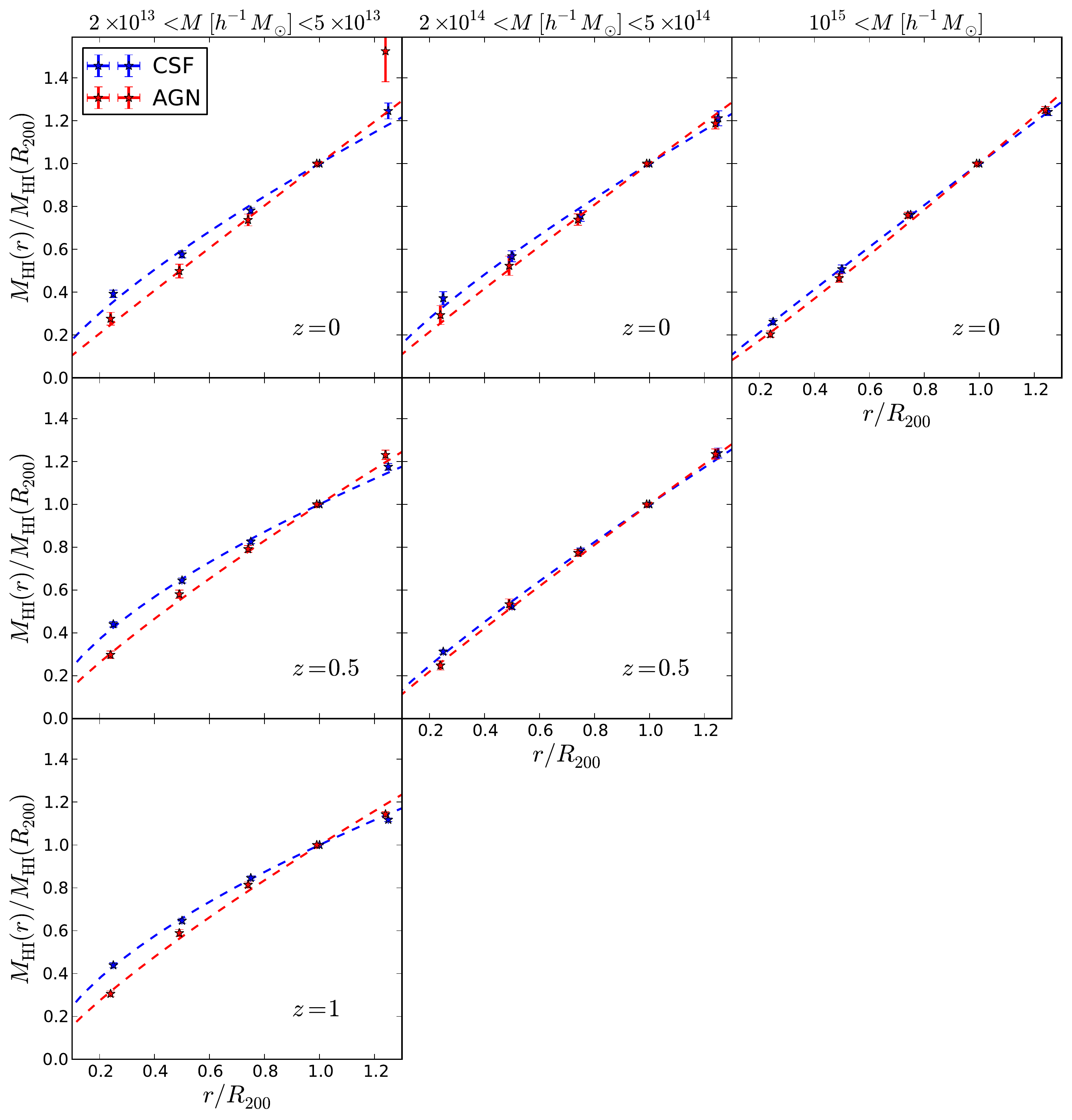}}
\caption{Neutral hydrogen mass within a radius $r$, normalized by the HI mass within $R_{200}$, as a function of $r/R_{200}$ for halos in the mass range $M\in[2-5]\times10^{13}~h^{-1}M_\odot$ (left column), $M\in[2-5]\times10^{14}~h^{-1}M_\odot$ (middle column) and $M>10^{15}~h^{-1}M_\odot$ (right column) at $z=0$ (upper row), $z=0.5$ (middle row) and $z=1$ (bottom row) for the halos of the simulation CSF (blue) and AGN (red). For each halo we have computed the HI mass within 0.25, 0.50, 0.75, 1.00 and 1.25 times $R_{200}$, and the points with the errorbars display the average and dispersion around the mean of the results. For clearness, we have displaced the results of the AGN simulation by $\triangle(r/R_{200})=-0.01$. Dashed lines show a fit to the results using the functional form $M_{\rm HI}(r)/M_{\rm HI}(R_{200})=(r/R_{200})^\beta$. The best-fit values of $\beta$ are specified in Table \ref{tbl:M_HI_r}.} 
\label{fig:HI_profile}
\end{figure*}

We have also investigated the spatial distribution of neutral hydrogen within halos. In each simulation suite, we have selected halos of approximately the same mass, in order to isolate any dependence of the HI distribution on the halo mass. In particular, we have taken all halos with masses in three different ranges: $M\in[2-5]\times10^{13}~h^{-1}M_\odot$, $~M\in[2-5]\times10^{14 }~h^{-1}M_\odot$ and $M>10^{15}~h^{-1}M_\odot$. Next, for each halo in a given mass range we have computed the HI mass within a radius equal to 1/4, 1/2, 3/4 and 5/4 times $R_{200}$. In Fig. \ref{fig:HI_profile} we show the average and dispersion around the mean of the HI mass within those radii normalized to the HI mass within $R_{200}$ as a function of $r/R_{200}$ for the three different mass ranges and for the two different simulations.

By construction, the point at $r/R_{200}=1$ has a value of $M_{\rm HI}(r|M,z)/M_{\rm HI}(R_{200}|M,z)$ equal\footnote{The notation $M_{\rm HI}(r|M,z)$ denotes the HI mass within a radius $r$ of a halo of mass $M$ at redshift $z$.} to 1 and therefore its dispersion is 0, independently of the simulation setup, redshift and mass range. We find that the results can be well fitted by a law of the form $M_{\rm HI}(r|M,z)/M_{\rm HI}(R_{200}|M,z)=(r/R_{200})^\beta$. In Table \ref{tbl:M_HI_r} we show the best-fit value of $\beta$ for the three different mass bins and the two simulations at different redshifts. The dashed lines of Fig. \ref{fig:HI_profile} display those fits at $z=0$ (upper row), $z=0.5$ (middle row) and $z=1$ (bottom row). We notice that the fits only use the points with $r<R_{200}$, i.e. the data points with $r=5R_{200}/4$ are not included in the fit.

\begin{table*}
\centering{
{\renewcommand{\arraystretch}{1.1}
\resizebox{12cm}{!}{
\begin{tabular}{|c|c|c|cc|}
\cline{3-5}
& & \multicolumn{3}{|c|}{Mass range $[h^{-1}M_\odot]$} \\
\hline
Simulation & z & ~~$[2-5]\times10^{13}$~~ & $[2-5]\times10^{14}$ & $>10^{15}$ \\
\hline
\multirow{3}{*}{CSF} & 0 &$0.75\pm0.05$ & $0.80\pm0.06$ & $0.966\pm0.007$\\
                                 & 0.5 &$0.62\pm0.02$ & $0.87\pm0.03$ & $-$\\
	                          & 1 &$0.60\pm0.01$ & $-$ & $-$\\

\hline
\multirow{3}{*}{AGN} & 0 & $0.98\pm0.04$ & $0.96\pm0.05$ & $1.09\pm0.05$ \\
                                 & 0.5 &$0.83\pm0.03$ & $0.94\pm0.04$ & $-$\\
	                          & 1 &$0.81\pm0.04$ & $-$ & $-$\\

\hline
\end{tabular}}}}
\caption{Best-fit value of $\beta$ ($M_{\rm HI}(r|M,z)/M_{\rm HI}(R_{200}|M,z)=(r/R_{200})^\beta$) for the two different simulations at redshifts $z=0, ~0.5$ and 1. Entries with $-$ indicate that there are not enough halos in that redshift and mass range to properly derive the value of $\beta$.}\vspace{-1.5em}
\label{tbl:M_HI_r}
\end{table*}

Our results point out that for the simulation CSF the value of $\beta$, which measures the steepness of the function $M_{\rm HI}(r|M,z)$, increases with the halo mass, while for the simulation AGN the value of $\beta$ is compatible, within $2\sigma$, with 1 for all mass ranges.

\subsection{Impact of AGN feedback}
\label{subsec:AGN_feedback}

Now we study the impact that AGN feedback induces on the mass and spatial distribution of neutral hydrogen in galaxy groups and clusters. As expected, when AGN feedback is turned on, the mass in neutral hydrogen of a given halo decreases on average (see Fig. \ref{fig:M_HI}). This happens because AGN feedback injects energy to the gas, increasing its temperature and therefore avoiding the formation of neutral hydrogen. This can be seen more clearly in the upper panel of Fig. \ref{fig:HI_correlations} where we show, for each halo in the CSF and AGN simulations, the mass in gas versus the mass in neutral hydrogen. It can be seen that the mass in gas does not change much by switching on or off AGN feedback (see the halo-by-halo comparison below), while halos in the CSF simulations contain a larger HI mass because the gas in the appropriate conditions to host HI is hotter when AGN feedback is on. By fitting the results to a law of the form $M_{\rm HI}(M,z)=e^\gamma M^\alpha$ we find that AGN feedback does not change much the slope of the function $M_{\rm HI}(M,z)$ ($\alpha=0.78$ for CSF and $\alpha=0.75$ for AGN) but its effect is mainly to shift the overall amplitude of that function ($\gamma=0.7$ for CSF versus $\gamma=0.9$ for AGN).

We notice that AGN feedback also affects the halo mass function \citep{Cui_2014, Velliscig_2014,Khandai_2014}. In particular, it is expected that halos in simulations with AGN feedback would be smaller and lighter than those found in simulations with no AGN feedback. Thus, one may wonder whether the HI mass deficit in halos where AGN is on, with respect to the same halos where AGN is off, is just due to the fact that those halos are smaller, i.e. simply because there is less gas in halos when AGN feedback is switched on. In order to answer this question we have performed a halo-by-halo comparison between the simulations CSF and AGN. In order to match a halo in the simulation CSF with the corresponding one in the simulation AGN we require the distance among their centers to be smaller than $15\%$ of the halo radius and, in any case, that the distance is lower than 150 comoving $h^{-1}$kpc. 

\begin{figure*}
\centerline{\includegraphics[width=1\textwidth]{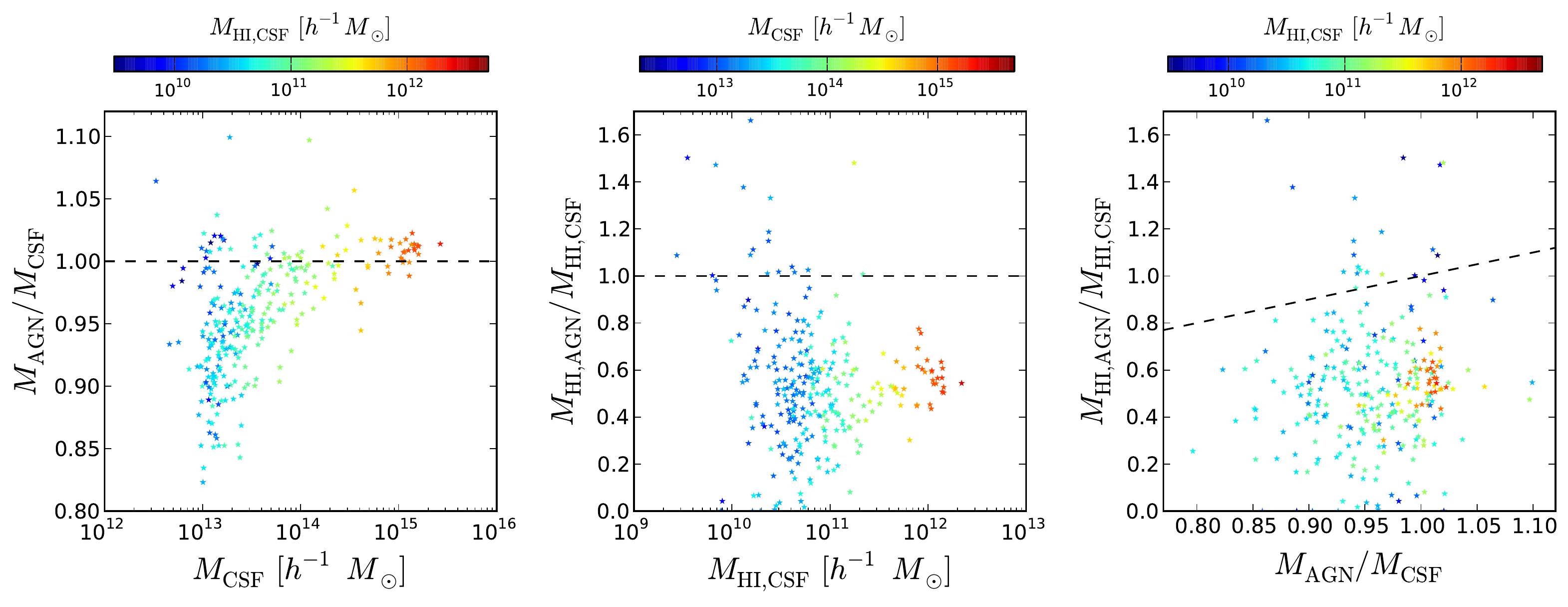}}
\caption{Impact of AGN feedback on the mass and HI content of groups and clusters at $z=0$. We carry out a halo-by-halo comparison among the simulations CSF and AGN. \textit{Left (middle):} Ratio between the mass (HI mass) of corresponding halos in the simulations AGN and CSF versus the mass (HI mass) of the CSF halo. The dashed horizontal line represents the case where the masses (HI masses) of the AGN and CSF halos are the same. \textit{Right:} Ratio between the HI masses of corresponding halos in the simulations AGN and CSF versus the ratio of their halo masses. The dashed line in that panel represents the curve $M_{\rm HI,AGN}/M_{\rm HI,CSF}=M_{\rm AGN}/M_{\rm CSF}$. The color of each point indicates the HI mass of the CSF halo (left and right panels) and its total mass (middle panel).}
\label{fig:AGN_feedback}
\end{figure*}

In the left panel of Fig. \ref{fig:AGN_feedback} we plot the ratio between the mass of a given halo in the AGN simulation to the mass of the same halo in the simulation CSF as a function of the CSF halo mass. Our results point out that, on average, AGN feedback reduces the total mass of galaxy groups by $\sim5-10\%$, while in clusters its effect is less important: halos more massive than $\sim10^{14.5}~h^{-1}M_\odot$ have almost the same total mass, independently on whether AGN feedback is switched on or off. The reason of this behavior is that AGN feedback is more effective removing baryons in low-mass halos than in the most massive ones.

The middle panel of Fig. \ref{fig:AGN_feedback} displays the ratio between the neutral hydrogen mass of corresponding halos of the simulations CSF and AGN as a function of the HI mass of the halo in the CSF simulation. The color of each point represents the total mass of the CSF halo. We find that AGN feedback can dramatically change the neutral hydrogen content of groups and clusters of galaxies. While in terms of total mass, AGN feedback can decrease the mass content by up to $~20\%$, when focusing on the HI content, AGN feedback can suppress its abundance by more than $95\%$. We note however that we have found a few halos where the HI content increases when AGN feedback is on (see the points above the dashed horizontal line in the middle panel of Fig. \ref{fig:AGN_feedback}).

Finally, in the right panel of Fig. \ref{fig:AGN_feedback} we show the ratio between the total masses versus the ratio of the HI masses for corresponding halos in the CSF and AGN simulations. As can be seen in that plot, the majority of points fall bellow the $y=x$ line, demonstrating that AGN feedback is more efficient removing HI gas than matter. In other words, the HI deficit we find in halos where AGN feedback is switched on, in comparison to the same halos with no AGN feedback, does not arise because those halos are smaller, but it is mainly due to effects induced by AGN feedback.

Regarding the impact that AGN feedback induces on the spatial distribution of neutral hydrogen within halos we can see from Fig. \ref{fig:HI_profile} that the steepness of the $M_{\rm HI}(r|M,z)/M_{\rm HI}(R_{200}|M,z)$ ratio is higher in the simulations where AGN feedback is on. This means that the fraction of HI residing in the inner regions of the halos is higher when AGN feedback is not active. This effect increases with decreasing halo mass, pointing out that AGN feedback impacts more strongly on the regions near the halo center, especially in the group regime.

\subsection{Redshift evolution}
\label{subsec:redshift_evolution}

In this subsection we investigate the time dependence of the functions $M_{\rm HI}(M,z)$ and $M_{\rm HI}(r|M,z)$. 

\begin{figure*}
\centerline{\includegraphics[width=1.0\textwidth]{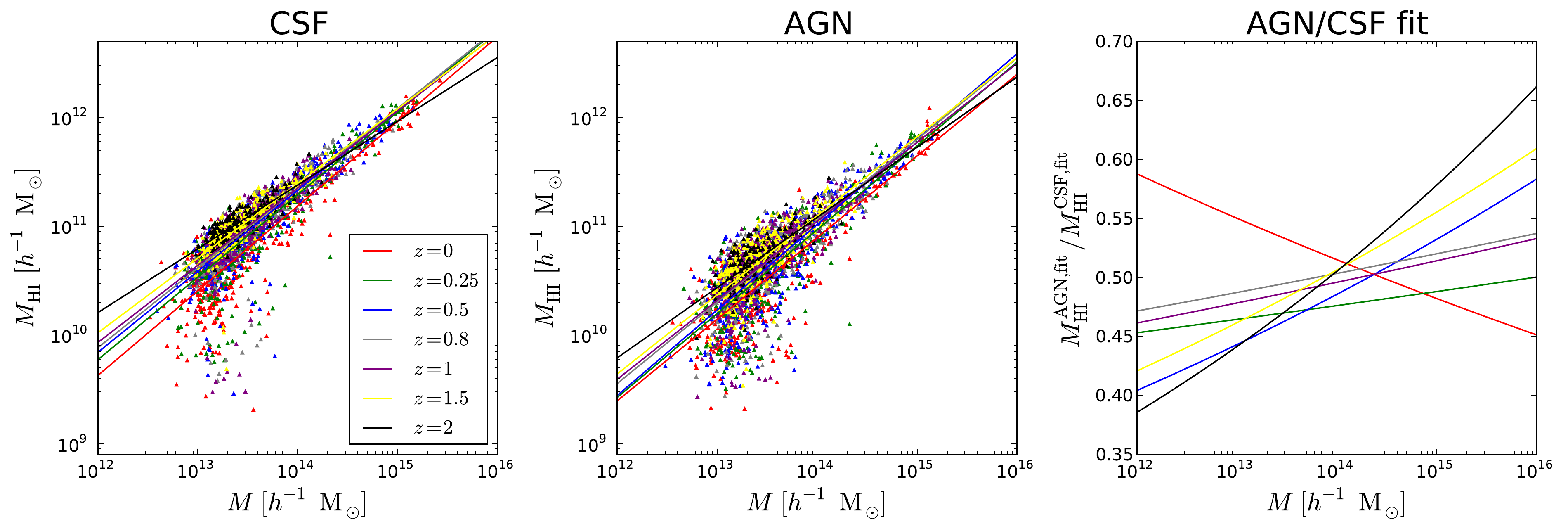}}
\caption{Redshift evolution of the $M_{\rm HI}(M,z)$ function for the simulation without and with AGN feedback, left and middle panel, respectively. Each point represents the HI mass inside $R_{200}$ of a halo of mass $M$, with the color of the point indicating the redshift (see legend). For each redshift, we fit the results to a law of the form $M_{\rm HI}(M,z)=e^\gamma M^\alpha$. The solid lines represent the best fits at the different redshifts. The right panel shows the ratio between the fit to the $M_{\rm HI}(M,z)$ function in the AGN and CSF simulations, as a function of the halo mass, at the different redshifts.}
\label{fig:redshift_evolution}
\end{figure*}

For each halo of both simulations with and without AGN feedback, at redshifts $z=0,~0.25,~0.5, ~0.8, ~1,~1.5$ and $2$ we have computed the neutral hydrogen mass hosted by the halo within $R_{200}$. In Fig. \ref{fig:redshift_evolution} we show the results for the CSF (left) and AGN (middle) simulations. 
We find that our results can be fitted by the same law we discussed in subsection \ref{subsec:HI_clusters}, $M_{\rm HI}(M,z)=e^\gamma M^\alpha$, i.e. a linear relation between the logarithms of $M_{\rm HI}$ and $M$. In Table \ref{tbl:M_HI} we show the best fit values for $\alpha$ and $\gamma$ for the two different simulations at the different redshifts. We also show in Fig. \ref{fig:redshift_evolution} with solid colored lines the fits.

For the simulations with no AGN feedback, we find that the slope of the $M_{\rm HI}-M$ relationship, given by $\alpha$, decreases with redshift, pointing out that the relative differences in the HI content between halos of different masses are smaller at high redshift than at low redshift. Moreover, the overall normalization of the $M_{\rm HI}(M,z)$ function, given by $\gamma$, increases its value with redshift. This means that, for halos with masses lower than $\sim10^{14}~h^{-1}M_\odot$, the neutral hydrogen content of halos with a fixed mass, increases with redshift, being the effect more pronounced in halos with low masses.

Our results for the simulations with AGN feedback switched on point out that $\alpha$ increases with redshift until $z\sim0.5$ while at higher redshifts it decreases its value. $\gamma$, on the other hand, decreases its value up to $z\sim0.5$ and increases its value at higher redshifts. We find that a single fit to all results between redshift 0 and redshift 1.5 is a very good description of the data, and we show in Table \ref{tbl:M_HI} the best fit values. 

The right panel of Fig. \ref{fig:redshift_evolution} displays the ratio between the fit to the $M_{\rm HI}(M,z)$ function in the simulations with AGN feedback on and off, at different redshifts. We find that for all masses and redshifts, the ratio is always below 1, demonstrating once again the effect of AGN feedback reducing the amount of neutral hydrogen. It is interesting to point out that AGN feedback tends to suppress the HI mass by roughly $50\%$, with a very weak dependence on redshift and halo mass. Indeed, the ratio $M_{\rm HI}^{\rm AGN}/M_{\rm HI}^{\rm CSF}$ is very stable around 0.5 while varying the halo mass by four orders of magnitude and from redshift $z=0$ to $z=2$. We notice that the quantity plotted in that panel is just $e^{\gamma_{\rm AGN}}M^{\alpha_{\rm AGN}}/(e^{\gamma_{\rm CSF}}M^{\alpha_{\rm CSF}})$, where AGN and CSF stand for the value of $\alpha$ and $\gamma$ for that simulation (see Table \ref{tbl:M_HI}). The mass dependence of that quantity is therefore given by $M^{\gamma_{\rm AGN}-\gamma_{\rm CSF}}$. At $z>0$ $\gamma_{\rm AGN}>\gamma_{\rm CSF}$ and thus the ratio increases with mass, while at $z=0$ the opposite situation takes place, inducing a change of slope as can be seen in the red line in that panel. We note that within $1\sigma$, the value of $\gamma_{\rm AGN}-\gamma_{\rm CSF}$ is also compatible with a positive number at $z=0$.


We have also investigated the time evolution of the spatial distribution of neutral hydrogen within halos. In the middle and bottom row of Fig. \ref{fig:HI_profile} we show the results of computing the ratio $M_{\rm HI}(r|M,z)/M_{\rm HI}(R_{200}|M,z)$, as a function of $r/R_{200}$ at redshift $z=0.5$ (middle) and $z=1$ (bottom). We notice that there are some situations in which the number of halos is either 0 or very small (for instance halos with masses larger than $10^{15}~h^{-1}M_\odot$ at $z\geqslant1$). In these cases we do not show the results.

We find that at redshifts higher than $z=0$, the fraction of the total neutral hydrogen that resides in a given halo is higher in the inner regions in simulations where AGN feedback is switched off. The same conclusions hold at $z=0$ (see subsection \ref{subsec:HI_clusters}). At $z=0.5$ we also observe the same trend with halo mass that we find at $z=0$: the differences between the results in simulations with and without AGN feedback decrease with increasing halo mass. 

As with the results at $z=0$, we find that a very simple law, of the form $M_{\rm HI}(r|M,z)/M_{\rm HI}(R_{200}|M,z)=(r/R_{200})^\beta$ can reproduce very well our measurements. In Table \ref{tbl:M_HI_r} we show the best-fit values of the parameter $\beta$ for different halo mass ranges and redshifts. Our results point out that for groups of galaxies ($[2-5]\times10^{13}~h^{-1}M_\odot$) the value of $\beta$ decreases with redshift, both for simulations with AGN feedback on or off. This means that the fraction of HI that it is located in the inner regions of a halo with a fixed mass, increases with redshift (at least until $z=1$). At $z=0.5$, we find that the value of $\beta$ increases with the halo mass, independently of the simulation used; a trend that we already found at $z=0$. Finally, the redshift evolution of $\beta$ for galaxy clusters is less evident, with $\beta$ increasing for the CSF simulations and decreasing for the AGN simulations. Notice however that given the error in the fits, the results at the two different redshifts are compatible at $1\sigma$.

\section{Comparison with theoretical models and implications for intensity mapping}
\label{sec:Bagla}

In this section we compare our results against the predictions of the \citet{Bagla_2010} model and study the implications for 21cm intensity mapping.

As discussed in the introduction, the shape and amplitude of the function $M_{\rm HI}(M,z)$ is of primary importance for future surveys that aim at putting constraints on the cosmological parameters using intensity mapping. We now compare our findings with the theoretical model of \cite{Bagla_2010}, which has been commonly used in the literature to perform forecasts \citep{Camera_2013, Bull_2015, Villaescusa-Navarro_2014a, Villaescusa-Navarro_2014b, Carucci_2015, Villaescusa-Navarro_2015a} and to create mock 21cm intensity mapping maps \citep{Seehars_2015}.

\citet{Bagla_2010} proposed a functional form for the $M_{\rm HI}(M,z)$ function as follows
\begin{equation}  
M_{\rm HI}(M,z) = \left\{ 
  \begin{array}{l l}	
  
    f_3(z)\frac{M}{1+M/M_{\rm max}(z)} & \quad \text{if $M_{\rm min}(z)\leqslant M$}
    \\
    0 & \quad \text{otherwise,}\\
  \end{array} \right.
\label{M_HI_Bagla3}
\end{equation} 
where the values of the free parameters $M_{\rm min}(z)$ and $M_{\rm max}(z)$ are taken in correspondence to the masses of halos with circular velocities equal to $v_{\rm min}=30$ km/s and $v_{\rm max}=200$ km/s at redshift $z$, respectively. $f_3(z)$ is also a free parameter, and its value is chosen to reproduce the HI density parameter $\Omega_{\rm HI}(z)$. 

With a dashed magenta line, we show in Fig. \ref{fig:M_HI} the prediction of the \citet{Bagla_2010} model for the function $M_{\rm HI}(M,z)$ at $z=0$. We find that the Bagla model underestimates the neutral hydrogen content of both clusters and groups of galaxies. The deviations are however more dramatic in galaxy clusters (for halos of $10^{15}~h^{-1}M_\odot$ differences are of two orders of magnitude), with the Bagla model predicting that those halos should host about the same HI mass while our results indicate that the neutral hydrogen mass increases with the mass of the host halo. Notice that the Bagla model also underpredicts, by almost one order of magnitude, the lower limits on the HI masses of the Sausage and Virgo clusters (see subsection \ref{subsec:HI_clusters}). By extrapolating our results to smaller halos, we find that the prediction of the Bagla model agrees pretty well with our fitting formula, for halos with masses lower than $\sim10^{12}~h^{-1}M_\odot$ when AGN feedback is switched off, while when it is switched on the Bagla model overpredicts the neutral hydrogen content of those halos.

Notice that we are restricting our analysis to the model 3 of \citet{Bagla_2010}, which physically is the one best motivated among the three different models proposed by those authors. In their model 1, \citet{Bagla_2010} proposed that halos with circular velocities larger than $200$ km/s ($\sim2\times10^{12}~h^{-1}M_\odot$ at $z=0$) do not host any HI at all. This is clearly in tension with our results but also with the lower limits on the HI mass of the Sausage and Virgo clusters. The model 2 of \citet{Bagla_2010} lies in between models 1 and 3, and therefore the discrepancies among our results and the observational limits will be larger than those obtained by employing the model 3. We therefore concentrate our analysis in the model 3 of \citet{Bagla_2010}.

\begin{figure*}
\centerline{\includegraphics[width=0.9\textwidth]{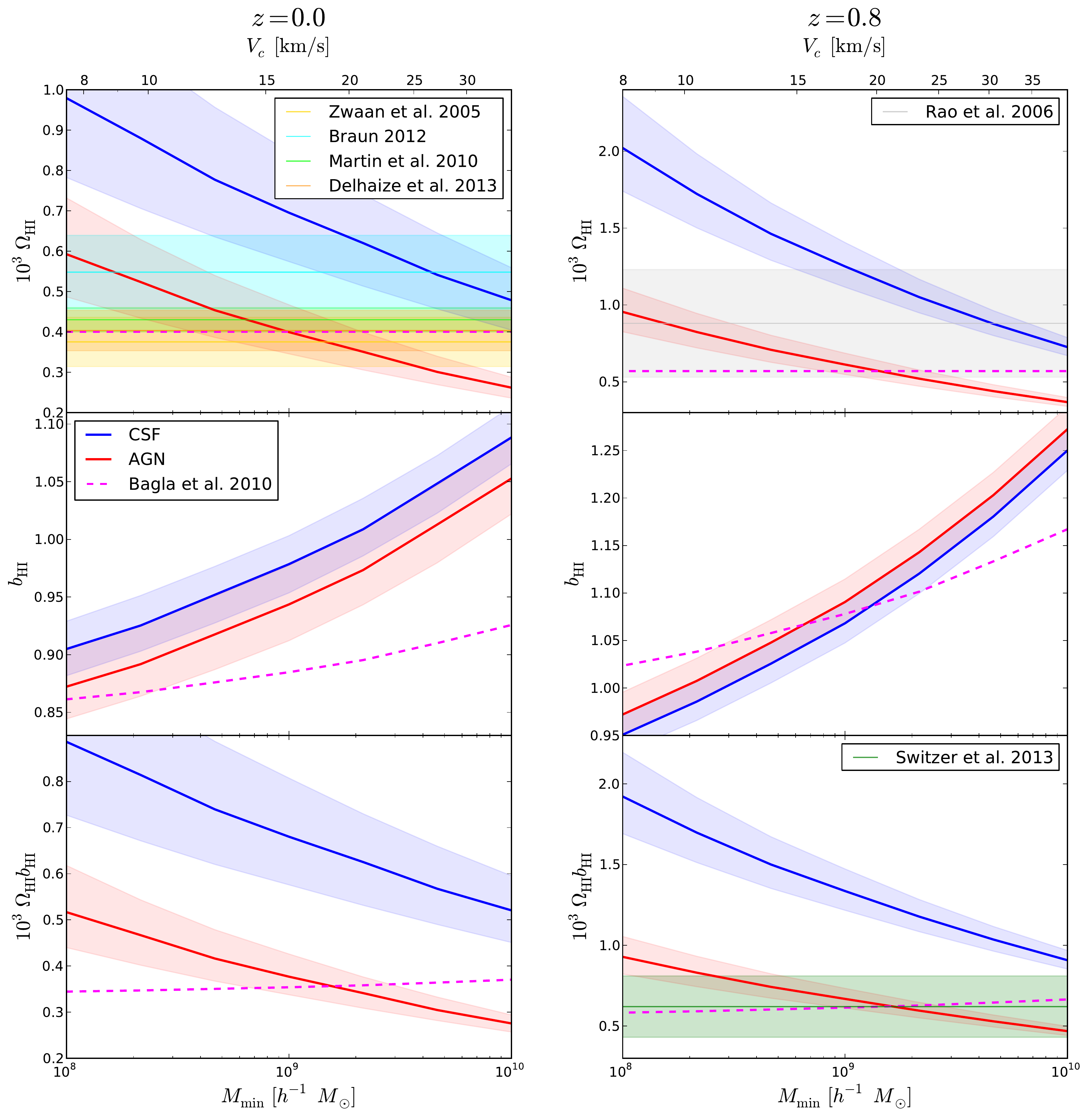}}
\caption{We use the functions $M_{\rm HI}(M)$ from the \citet{Bagla_2010} model (purple dashed line) and from the fit to the simulations (red and blue solid lines; accompanying colored areas indicate the variation of the results according to the fitting errors) to compute the value of $\Omega_{\rm HI}$ (upper panel), $b_{\rm HI}$ (lower panel) and $\Omega_{\rm HI}b_{\rm HI}$ at $z=0$ (left column) and $z=0.8$ (right column). We extrapolate the fits to halo masses lower than the simulation resolution limit assuming that halos with masses lower than $M_{\rm min}$ do not host any HI. The plots show the results as a function of $M_{\rm min}$. The colored bands in the upper-left panel display the value of $\Omega_{\rm HI}(z\cong0)$ from observations \citep{Zwaan_2005,Braun_2012,Martin_2010,Delhaize_2013} and in the upper-left panel the value of $\Omega_{\rm HI}(z\cong0.53\pm0.38)$ from \citet{Rao_2006}. The green band in the bottom-right panel shows the measurement of $\Omega_{\rm HI}b_{\rm HI}$ at $z\sim0.8$ from \citet{Switzer_2013}.}
\label{fig:HI_bias}
\end{figure*}

We notice that the fact that the Bagla model underestimates the neutral hydrogen mass in massive halos has important consequences with what respects the spatial distribution of HI, as we shall see now. As we discussed in the introduction, two of the most important quantities for 21cm intensity mapping are average neutral hydrogen density in units of the Universe critical density, $\Omega_{\rm HI}$, and the bias of that distribution with respect to the one of total matter, $b_{\rm HI}$. These quantities can be easily computed once the shape and amplitude of the function $M_{\rm HI}(M)$ is known. In Fig. \ref{fig:HI_bias} we show with a dashed magenta line the values of $\Omega_{\rm HI}$, $b_{\rm HI}$ and $b_{\rm HI}\Omega_{\rm HI}$ that we obtain by using the Bagla model as a function of the minimum mass of halos that host HI, $M_{\rm min}$ at $z=0$ (left column) and $z=0.8$ (right column). 

As we have discussed above, the Bagla model has one free parameter, $f_3(z)$, whose value is chosen to reproduce the value of $\Omega_{\rm HI}(z)$. In our case, we follow \citet{Crighton_2015} and we assume $\Omega_{\rm HI}(z)=4\times10^{-4}(1+z)^{0.6}$, in excellent agreement with the observational measurements at $z=0$ by \citet{Zwaan_2005,Braun_2012,Martin_2010,Delhaize_2013} and at $z\sim0.8$ by \citet{Rao_2006}. We show these observational measurements as colored bands in the upper panels of Fig. \ref{fig:HI_bias}. Thus, by construction, the value of $\Omega_{\rm HI}$ does not depend on $M_{\rm min}$ when using the Bagla model. On the other hand, the value of $b_{\rm HI}$ does depend on $M_{\rm min}$, as shown in the middle panels of Fig. \ref{fig:HI_bias}. We find however that the dependence of $b_{\rm HI}$ on $M_{\rm min}$ is very weak, with $b_{\rm HI}$ ranging only from $\sim0.85$ to $\sim0.92$ at $z=0$ and from $\sim1.03$ to $\sim1.15$ at $z=0.8$ when $M_{\rm min}$ is varied over two orders of magnitude. 

The bottom panels of Fig. \ref{fig:HI_bias} show the value of $\Omega_{\rm HI}b_{\rm HI}$ as a function of $M_{\rm min}$. We remind the reader that the relevant quantity for 21cm intensity mapping is the product $\Omega_{\rm HI}b_{\rm HI}$ (see Eq. \ref{eq:P_21cm}) and not the values of $\Omega_{\rm HI}$ and $b_{\rm HI}$ separately. Given the fact that $\Omega_{\rm HI}(z)$ is fixed in the Bagla model, and that $b_{\rm HI}(z)$ barely changes with $M_{\rm min}$ it is not surprising that the quantity $\Omega_{\rm HI}b_{\rm HI}$ exhibits such a weak dependence with $M_{\rm min}$. At $z\sim0.8$, 21cm intensity mapping observations have determined the value of $\Omega_{\rm HI}b_{\rm HI}$ to be $6.2^{+2.3}_{-1.5}\times10^{-4}$ \citep{Chang_2010,Masui_2013,Switzer_2013}. In the bottom-left panel of Fig. \ref{fig:HI_bias} we show with a colored green band those results. We find that Bagla model reproduces very well those observations, as was already pointed out in \citet{Padmanabhan_2015}. Unfortunately, no measurements of $\Omega_{\rm HI}b_{\rm HI}$ are available at other redshifts. 

We have also used the fitting function which reproduces the results of our simulations, extrapolating it to halo masses below the resolution limit of our simulations, considering, as in the Bagla model, that only halos above $M_{\rm min}$ host HI. In other words, we have modeled the function $M_{\rm HI}(M,z)$, according to the results of our simulations as
\begin{equation}  
M_{\rm HI}(M,z) = \left\{ 
  \begin{array}{l l}	
  
    e^\gamma M^\alpha & \quad \text{if $M_{\rm min}(z)\leqslant M$}
    \\
    0 & \quad \text{otherwise,}\\
  \end{array} \right.
\label{M_HI_fit}
\end{equation} 
where the values of the parameters $\alpha$ and $\gamma$ are given in Table \ref{tbl:M_HI}. We take the values of $\alpha$ and $\gamma$ at $z=0$ and $z=0.8$ from Table \ref{tbl:M_HI}. In Fig. \ref{fig:HI_bias} we show the value of $\Omega_{\rm HI}$, $b_{\rm HI}$ and $\Omega_{\rm HI}b_{\rm HI}$ as a function of $M_{\rm min}$, that we obtain using the above $M_{\rm HI}(M,z)$ function for both the CSF and AGN simulations. We also show with colored bands around the mean value, the variation of the results arising from the $1\sigma$ uncertainty in the value of $\alpha$ and $\gamma$ from Table \ref{tbl:M_HI}.

As expected, given the fact that AGN suppresses the HI content of halos, for a fixed value of $M_{\rm min}$, the value of $\Omega_{\rm HI}$ is always lower when using the $M_{\rm HI}(M,z)$ function from the AGN simulations than from the CSF simulations. We find that in order to reproduce the value of $\Omega_{\rm HI}(z)$ from observations, halos with circular velocities higher than $\sim20~{\rm km/s}$ at $z=0$ and $\sim25~{\rm km/s}$ at $z=0.8$ are required to host HI when AGN feedback is turned off. On the other hand, when AGN feedback is on, we conclude that halos with circular velocities higher than $\sim25~{\rm km/s}$, at both $z=0$ and $z=0.8$, should host HI. Notice that the standard Bagla model assumes that only halos with $V_c>30~{\rm km/s}$ contain HI.

The value of the HI bias, that we computed using Eq. \ref{eq:bias_HI} and making use of \citet{ST} and \citet{SMT} models for the halo mass function and halo bias, respectively, point out that the HI bias is higher in the CSF model at $z=0$, in comparison with the AGN model, for all values of $M_{\rm min}$, whereas the opposite situation takes place at $z=0.8$. We notice that the value of the HI bias does not depend on $\Omega_{\rm HI}$, but just on the way the HI is distributed among the different halos. This implies that for the AGN and CSF models, for which we use $M_{\rm HI}(M,z)=e^\gamma M^\alpha$, the value of the HI bias only depends on $\alpha$. Thus, the reason of why the HI bias is higher in the CSF model at $z=0$, in comparison with the AGN model, is simply because $\alpha$ has a higher value in the CSF model at $z=0$. The same arguments holds at $z=0.8$, where the CSF model has a lower value of $\alpha$ than the AGN model. 

By comparing the results of the AGN and CSF models with those of the Bagla model we find that at $z=0$, and for the values of $M_{\rm min}$ considered, the HI bias is always higher in the AGN and CSF models. The reason of this behavior is again the distribution of HI among the different halos: in the CSF and AGN models a significant amount of HI is placed on clusters of galaxies, which are very biased objects and therefore it is not surprising that the value of the HI bias increases in those models. On the other hand, at $z=0.8$ we find that the value of the HI bias in the AGN and CSF models is higher than the one of the Bagla model only if $M_{\rm min}$ is higher than $\sim10^{9}~h^{-1}M_\odot$. This is again a consequence of the distribution of HI among the different halos. 

Finally, in the bottom panel of Fig. \ref{fig:HI_bias} we show the value of $\Omega_{\rm HI}b_{\rm HI}$ as a function of $M_{\rm min}$ for the AGN and CSF models. We find that the CSF model predicts a value of $\Omega_{\rm HI}b_{\rm HI}$ significantly higher than these of the AGN and Bagla models, in disagreement with the observational measurements at $z=0.8$ by \citet{Switzer_2013}. On the other hand, the AGN model is capable of reproducing those measurements extremely well for almost all $M_{\rm min}$ masses. It is interesting to notice that for a value of $M_{\rm min}\sim25~{\rm km/s}$, for which the AGN model is capable of reproducing the observed value of $\Omega_{\rm HI}$ at both $z=0$ and $z=0.8$, the value of $\Omega_{\rm HI}b_{\rm HI}$ is very similar among the AGN and Bagla models. 

We now discuss the robustness of our results. As we saw in subsection \ref{subsec:HI_clusters}, our results overpredict the HI mass in the most massive halos by a factor of $3-4$. This opens different possibilities. The first one is that our simulations predict correctly the HI mass in groups and small clusters but overpredict the neutral hydrogen mass in the most massive halos. In that case, the quoted values of $\Omega_{\rm HI}$ and $b_{\rm HI}$ will barely changed since the amount of HI in those halos is small\footnote{We have explicitly checked this by repeating the analysis using the function \begin{equation}  
M_{\rm HI}(M,z) = \left\{ 
  \begin{array}{l l}	
    e^\gamma M_{\rm max}^\alpha & \quad \text{if $M_{\rm max}\leqslant M$}  
  \\
    e^\gamma M^\alpha & \quad \text{if $M_{\rm min}(z)\leqslant M\leqslant M_{\rm max}$}
    \\
    0 & \quad \text{otherwise,}\\
  \end{array} \right.
\label{M_HI_fit}
\end{equation} with $M_{\rm max}=5\times10^{14}~h^{-1}M_\odot$ and we found that differences are very small. Notice that this function will be able to explain the HI masses of the Virgo and Sausage clusters.}. The second possibility is that our simulations overpredict the value of $M_{\rm HI}(M,z)$ for all halo masses. In that case, the value of $\gamma$ will decrease keeping fixed the value of $\alpha$. Under those circumstances, the value of $b_{\rm HI}$ will remain unchanged but the value of both  $\Omega_{\rm HI}$ and $\Omega_{\rm HI}b_{\rm HI}$ will decrease by a factor $3-4$, putting the model in difficulties to reproduce the observational results. Finally, it could happen that our simulations overpredict the HI mass in the most massive clusters but underpredict the HI mass in small groups (due to resolution limitations and because numerical contamination is stronger in clusters than in groups). In that situation the value of $\alpha$ will decrease, affecting both the value of the HI bias, that will decrease, and the value of $\Omega_{\rm HI}$, that will increase. 

\section{Summary and discussion}
\label{sec:conclusions}

Future radio-surveys will sample the large scale structure of the Universe by detecting the 21cm emission from cosmic neutral hydrogen. Two different techniques can be employed for cosmological studies: 1) perform a HI galaxy survey, where individual galaxies are detected and 2) use the intensity mapping technique to carry out a low angular resolution survey where the integrated emission from many galaxies is measured. It has been recently pointed out that using an instrument like the future SKA, the constraints on the cosmological parameters will be much tighter if the intensity mapping technique is employed \citep[see][]{Bull_2015}.

In order to extract the maximum information from these surveys, an exquisite knowledge of the spatial distribution of neutral hydrogen is needed from the theoretical side. A key ingredient for doing cosmology with 21cm intensity mapping is the function $M_{\rm HI}(M,z)$, which represents the average mass in neutral hydrogen that host a halo of mass $M$ at redshift $z$. The reason is that once that function is known, one can compute the bias of the HI, the value of $\Omega_{\rm HI}(z)$ and $\delta T_b(z)$, and predict the shape and amplitude of the 21cm power spectrum on large, linear scales. 

It is thus very important to model that function as best as possible, in order to properly predict the bias and the amount of HI, which in the end will determine the expected signal-to-noise ratio of a given survey. The function $M_{\rm HI}(M,z)$ also carries very important astrophysical information, since neutral hydrogen represents an intermediate state between the highly ionized gas in the intergalactic medium and the dense molecular hydrogen that will end up forming stars. The amount of HI in halos of different mass is largely affected by the astrophysical processes that are undergoing in them, and therefore, we can use the function $M_{\rm HI}(M,z)$ to improve our understanding of the physical processes affecting the formation and evolution of galaxies.

In this paper we have studied the high-mass end of the $M_{\rm HI}(M,z)$ function using a set of zoom-in hydrodynamic simulations. Our simulations comprise two different sets, one in which AGN feedback is switched off (CSF) and another one which incorporates AGN feedback (AGN). All simulations incorporate metal-dependent radiative cooling, metal enrichment, supernova feedback and star formation. We have post-processed the output of the simulations to account for two critical processes needed to properly model the spatial distribution of HI: the HI self-shielding and the formation of molecular hydrogen. To correct for the former we use the fitting formula of \citet{Rahmati_2013} while for the latter we employ the KMT model \citep{Krumholz_2008, Krumholz_2009, McKee_2010}.

We find that the larger the mass of a halo the higher the neutral hydrogen mass it contains, a result that holds at all redshifts studied in this paper ($z\in[0-2]$) and for halos in the CSF and AGN simulations. We find that the results of the simulations can be well fitted by a law of the form: $M_{\rm HI}(M,z)=e^\gamma M^\alpha$. In Table \ref{tbl:M_HI} we show the best-fit values of $\gamma$ and $\alpha$ at different redshifts and for the two different simulation sets. 

We have also investigated the spatial distribution of neutral hydrogen within halos. In particular, we have studied the function $M_{\rm HI}(r|M,z)$, i.e. the function that gives the average HI mass within a radius $r$ for a halo of mass $M$ at redshift $z$. We find that for both halos in the CSF and AGN simulations the function $M_{\rm HI}(r|M,z)$ is well described by a law of the form, $M_{\rm HI}(r|M,z)/M_{\rm HI}(R_{200}|M,z)=(r/R_{200})^\beta$. Our results point out that the value of $\beta$ increases with the halo mass, implying that the fraction of HI residing in the inner regions of the halo decreases with the halo mass. We also observe a dependence of $\beta$ with redshift: for a fixed halo mass, the value of $\beta$ decreases with redshift. 

We have paid special attention to the effect that AGN feedback induces on the amount and spatial distribution of neutral hydrogen by comparing results among the CSF and AGN simulations. Regarding the function $M_{\rm HI}(M,z)$, we find that AGN feedback acts by decreasing the total amount of neutral hydrogen hosted by a given halo. Our results point out that the value of the slope of the function $M_{\rm HI}(M,z)$ is not significantly changed by AGN feedback, being its main effect to modify its overall normalization. We find that on average, AGN feedback decreases the neutral hydrogen content a halo host by $\sim50\%$, with a weak dependence on halo mass and redshift. 

AGN feedback also induces a shift in the value of the parameter $\beta$, meaning that the spatial distribution of neutral hydrogen within halos is also affected by that process. We find that AGN feedback reduces the fraction of HI that is located in the halo inner regions, i.e. AGN feedbacks tends to increase the value of $\beta$.

We have also performed a halo-by-halo comparison among the simulations CSF and AGN. We find that AGN feedbacks decreases both the HI mass and total halo mass. We then asked ourselves whether the deficit in HI we find in simulations with AGN feedback may just be due to the fact that those halos are simply smaller and therefore they host less HI. We find that AGN feedback is more effective reducing the amount of neutral hydrogen than the total mass. 

We have investigated the neutral hydrogen content in galaxies belonging to groups and clusters. We find that in groups of galaxies most of the overall HI mass resides within galaxies, while in clusters the fraction of the total HI mass made up by neutral hydrogen within galaxies is small. The reason is that in massive clusters gas can be more efficiently stripped from galaxies, mainly by the interaction with the hot, diffuse, intra-cluster medium. It is worth reminding that numerical effects, may create spurious blobs of HI that may bias our results. We have quantified the importance of these non-physical effects and we find that they are not important for groups, while for clusters they can contribute to the overall HI mass. We conclude that the simulated HI masses within clusters can only be trusted at this stage within a factor of 2.

We have compared our results for the $M_{\rm HI}(M,z)$ function against the prediction of the model by \citet{Bagla_2010}. We conclude that the \citet{Bagla_2010} model dramatically underpredicts the amount of neutral hydrogen that resides in clusters of galaxies (by more than 2 orders of magnitude for halos with masses of $10^{15}~h^{-1}M_\odot$), while in groups of galaxies the model and our results agree better, in particular for the simulation with AGN feedback on.

We derive consequences for 21cm intensity mapping by extrapolating our results for the $M_{\rm HI}(M,z)$ function below the mass resolution limit of our simulations. The fact that galaxy clusters host a much higher mass in neutral hydrogen than the one predicted by the Bagla model implies that, at $z=0$, the bias of the HI will be significantly higher than the one obtained by employing the Bagla model. We stress that the bias of the HI depends only on the slope of the $M_{\rm HI}(M,z)$ function and not on its overall normalization. On the other hand, the value of $\Omega_{\rm HI}(z)$ depends explicitly on the normalization of the function $M_{\rm HI}(M,z)$. We find that in order to reproduce the value of $\Omega_{\rm HI}(z)$ from observations, halos with circular velocities higher than $\sim 20~{\rm km/s}$, at $z=0$, and $\sim25~{\rm km/s}$ at $z=0.8$ must host HI when AGN feedback is off, while for the simulations where AGN feedback is on we conclude that halos with circular velocities higher than $\sim25~{\rm km/s}$ must host neutral hydrogen. We emphasize that the important quantity for 21cm intensity mapping is $\Omega_{\rm HI}b_{\rm HI}$, which determines the signal-to-noise ratio of the signal. We find that the CSF model predicts a much higher value of $\Omega_{\rm HI}b_{\rm HI}$ than the AGN and Bagla models, in tension with observations at $z=0.8$. On the other hand, the AGN and Bagla models are capable of reproducing those observations very well and at $z=0$, and assuming that only halos with circular velocities higher than $\sim25~{\rm km/s}$ host HI, both models predict a similar value of $\Omega_{\rm HI}b_{\rm HI}$. We therefore conclude that although the Bagla model underpredicts the mass of HI in galaxy clusters, it is a well calibrated model for 21cm intensity mapping. 

We emphasize that our AGN simulations overpredict the HI mass of the Virgo and Sausage cluster by a factor of $3-4$, and this may bias our conclusions for 21cm intensity mapping. It could happen that our simulations only overpredict the HI mass of the most massive clusters. In that case, the consequences for intensity mapping will be unchanged, since the HI mass in those halos is only a small fraction of the overall HI mass. On the other hand, if our simulations overpredict the HI mass of all halos, very low circular velocity halos will need to host HI in order to reproduce observations. It may also happen that our simulations overpredict the HI masses of very massive halos while they underpredict the HI masses of small groups. This situation may arise since numerical contamination is more important in clusters than in groups and due to the limited resolution of our simulations. In that case, the slope of the $M_{\rm HI}(M,z)$ function will decrease, impacting the value of the HI bias, that will decrease, and the value of $\Omega_{\rm HI}$, that will increase. We plan to use data from observations to study the $M_{\rm HI}(M,z)$ function and distinguish among the above different possibilities in a future paper.

Finally, we propose our own model for the $M_{\rm HI}(M,z)$ function which comes out from the results of our hydrodynamic simulations with AGN feedback and that by construction will be able to reproduce the observational constraints. We suggest to model the function $M_{\rm HI}(M,z)$ as
\begin{equation}  
M_{\rm HI}(M,z) = \left\{ 
  \begin{array}{l l}	
  
    e^{\gamma(z)} M^{\alpha(z)} & \quad \text{if $M_{\rm min}(z)\leqslant M$}
    \\
    0 & \quad \text{otherwise,}\\
  \end{array} \right.
\label{M_HI_Bagla3}
\end{equation} 
with $\alpha(z)=3/4$, in agreement with the results of our AGN simulations. $M_{\rm min}(z)$ represents the mass of the smallest halo that is capable of hosting HI at redshift $z$, and $\gamma(z)$ is the value of the overall normalization of the $M_{\rm HI}(M,z)$ function. Notice that we are assuming that effects as radiation from local sources and from the ICM and the limited resolution of our simulations only affect the value of $\gamma(z)$ and not the one of $\alpha(z)$, which may not be the case. In our model we will further assume that $M_{\rm min}(z)$ does not depend on redshift. 

The values of $M_{\rm min}$ and $\gamma(z)$ can be obtained by requiring that the function $M_{\rm HI}(M,z)$ reproduces observations of both $\Omega_{\rm HI}(z)$ and $b_{\rm HI}\Omega_{\rm HI}(z=0.8)$:
\begin{eqnarray}
\Omega_{\rm HI}(z) &=& \frac{e^{\gamma(z)}}{\rho_c^0}\int_{M_{\rm min}}^\infty n(M,z) M^{3/4}dM \\
\Omega_{\rm HI}b_{\rm HI}(z=0.8) &=& \frac{e^{\gamma(0.8)}}{\rho_c^0}\times\\
&&\int_{M_{\rm min}}^\infty n(M,0.8) b(M,0.8)M^{3/4}dM\nonumber\\
\end{eqnarray}
where $\rho_c^0$, $n(M,z)$ and $b(M,z)$ are the actual value of the Universe critical density, the halo mass function and the halo bias, respectively. The value of $\Omega_{\rm HI}b_{\rm HI}$ at $z=0.8$ has been measured in \citet{Chang_2010,Masui_2013,Switzer_2013} obtaining a value equal to $6.2^{+2.3}_{-1.5}\times10^{-4}$. The value of $\Omega_{\rm HI}(z)$ has been measured by different authors at different redshifts. A parametrization has been recently proposed parametrization by \citet{Crighton_2015}, $\Omega_{\rm HI}(z)=A(1+z)^\gamma$, with $A=(4.00\pm0.024)\times10^{-4}$ and $\gamma=0.60\pm0.05$.
Thus, our model for the $M_{\rm HI}(M,z)$ function will, by construction, reproduce the observations and at the same time the shape of that function will be in agreement with the results of our hydrodynamic simulations. 
We finally notice that since the contribution of spurious HI blobs could be more important in clusters than in groups, and since the limited resolution of our simulations may also impact on the HI mass of small groups, the actual value of $\alpha(z)$ might be lower than $3/4$ or alternatively, the $M_{\rm HI}(M,z)$ may exhibit a plateau on the high-mass end.

We conclude by pointing out some effects that may impact on our results. First of all, the resolution of our simulations\footnote{We notice that according to the resolution tests of \cite{Duffy_2012} and their derived criteria for convergence, the profiles, total masses and HI masses of the large majority of our halos should be converged against resolution.}does not allow us to follow the evolution of neutral hydrogen within small galaxies. The stability and robustness of our results should therefore be checked against resolution. We leave this for a future work. Secondly, the value of $\Omega_{\rm b}$ in our simulations simulations is $\sim20\%$ lower than the one found by Planck \citep[$\Omega_{\rm b}=0.049$,][]{Planck_2015}. We expect that if there is more gas in the Universe there will be more neutral hydrogen. Thirdly, we notice that in our analysis we have neglected the radiation from local sources and X-ray emission from the hot ICM. Accounting for these effects would require including a radiative-transfer description of the local radiation field that it is beyond the scope of the present paper. On the other hand, the mismatch between the positions of the HI blobs and the galaxies within clusters in our simulations may significantly decrease the importance of the radiation from local sources. 
And finally, a more profound and detailed analysis is required to identify the nature (physical/numerical) of the neutral hydrogen clouds outside galaxies we find in our simulations. We plan to investigate these points in a future paper.

\section*{ACKNOWLEDGEMENTS} 
We thank the referee, Alan Duffy, for his useful and constructive report. FVN thanks Andra Stroe, Rhys Taylor and Martha Haynes for useful discussions. FVN and MV are supported by the ERC Starting Grant ``cosmoIGM'' and partially supported by INFN IS PD51 ``INDARK''. SP, SB and GM are supported by the PRIN-INAF12 grant 'The Universe in a Box: Multi-scale Simulations of Cosmic Structures', the PRINMIUR 01278X4FL grant 'Evolution of Cosmic Baryons', and the INDARK INFN grant. SP acknowledges support by the Spanish Ministerio de Economia y Competitividad (MINECO, grants AYA2013-48226-C3-2-P) and the Generalitat Valenciana (grant GVACOMP2015-227). ER is supported by FP7-PEOPLE-2013-IIF (grant Agreement PIIF-GA-2013-627474) and NSF AST-1210973. KD acknowledges the support by the DFG Cluster of Excellence "Origin and Structure of the Universe". AMB is supported by the DFG Research Unit 1254 'Magnetisation of Interstellar and Intergalactic Media' and by the DFG Cluster of Excellence 'Origin and Structure of the Universe'. Simulations are carried out using Flux HCP Cluster at the University of Michigan efficiently supported by ARC-TS, and at CINECA (Italy), with CPU time assigned through ISCRA proposals and through an agreement with the University of Trieste. Data post-processing and storage has been done on the CINECA facility PICO, granted us thanks to our expression of interest. We acknowledge partial support from ``Consorzio per la Fisica - Trieste''.

\begin{appendix}

\section{Neutral hydrogen in galaxies and numerical artifacts}
\label{sec:HIMF}

In this Appendix we examine the neutral hydrogen content in galaxies, its overall contribution to the total HI mass in groups and clusters and the numerical problems our simulations may face.

A very interesting question to address would be: how does the HI mass function (HIMF) change as a function of the halo mass? In other words, how does the distribution of HI mass in galaxies vary with the mass of the  halo?

\begin{figure}
\centerline{\includegraphics[width=0.49\textwidth]{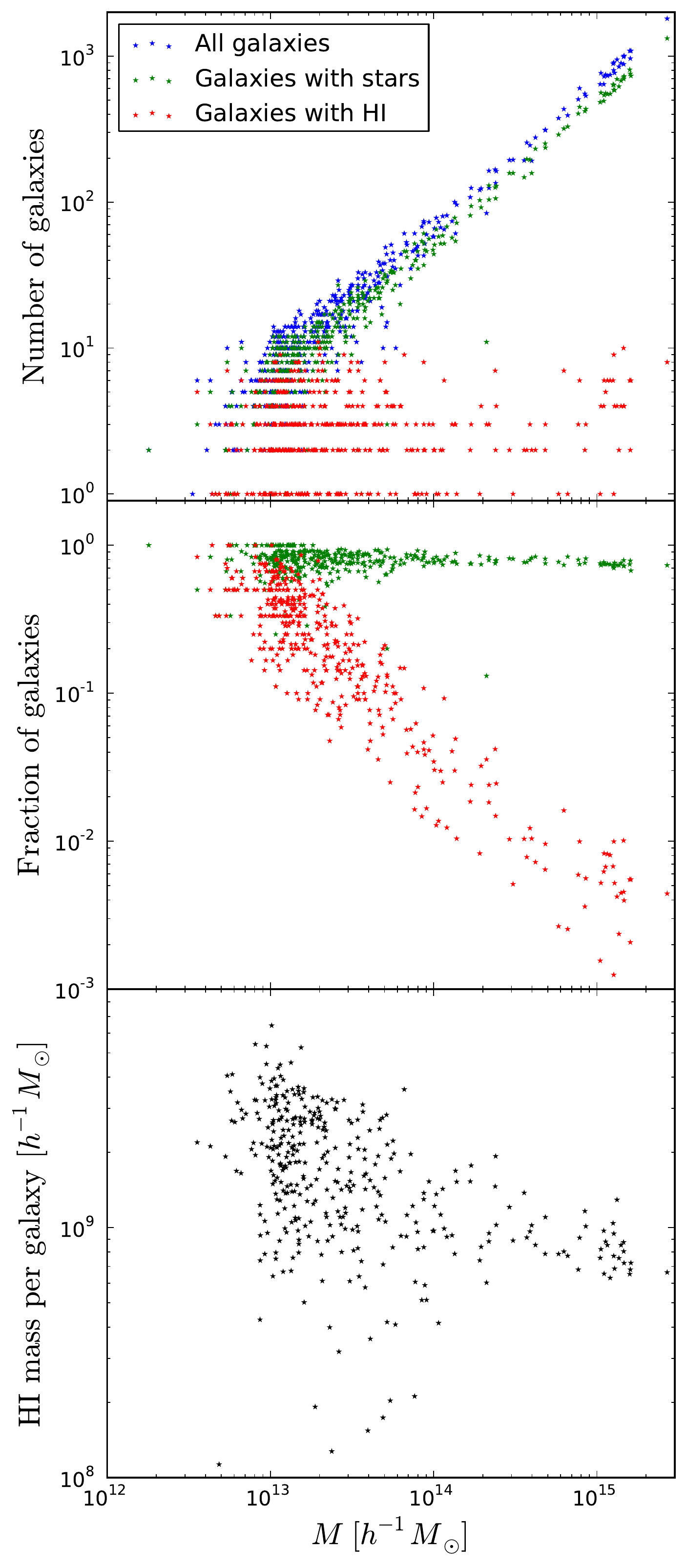}}
\caption{For each halo of the AGN simulations we compute the total number of subhalos (blue points), the number of subhalos with star particles (green points) and the number of subhalos with neutral hydrogen (red points) at $z=0$. The upper panel shows the different number of subhalos as a function of the halo mass, while the middle panel displays the fraction of subhalos hosting stars and neutral hydrogen. The bottom panel shows the average HI mass per galaxy, computed by dividing the halo HI mass by the total number of galaxies.}
\label{fig:Number_galaxies}
\end{figure}

In order to address that question, we have computed, for each subhalo, its stellar and neutral hydrogen mass. In the upper panel of Fig. \ref{fig:Number_galaxies} we show, for each halo in the simulations with AGN feedback, the total number of subhalos, together with the number of subhalos hosting stars and the number of subhalos containing HI.

As expected, we find that the total number of subhalos increases with the mass of the halo, a trend that is also followed by the number of subhalos hosting stars. We notice that, in general, the number of galaxies (i.e. subhalos containing stars) will be smaller than the total number of subhalos. This happens because {\sc SUBFIND} identifies subhalos up to the resolution limit (fixed in subhalos with at least 32 particles). Thus, very small subhalos are expected to not contain any star particle.

On the other hand, we find that the number of subhalos hosting neutral hydrogen does not increase with the halo mass, and that its number is never above $\sim10$ \citep[see also][for a similar analysis with observations]{Hess_2013}. This is mainly due to gas stripping by the ICM\footnote{We have also identified some cases in which gas stripping is produced by gravitational collision.}. This effect, originated by the interaction of the galaxy gas with the hot, diffuse, ICM it is expected to be more efficient removing gas of the galaxies residing in the most massive halos. In the middle panel of Fig. \ref{fig:Number_galaxies} we show the fraction of subhalos hosting HI as a function of the halo mass. Indeed, we find that this fraction quickly decreases with the mass of the halo, pointing out that gas stripping is more efficient in clusters than in groups. Notice that in groups of galaxies ($M_{\rm halo}\sim10^{13}~h^{-1}M_\odot$) the fraction of subhalos hosting HI is comparable to the fraction of subhalos with stars. 

\begin{figure}
\centerline{\includegraphics[width=0.5\textwidth]{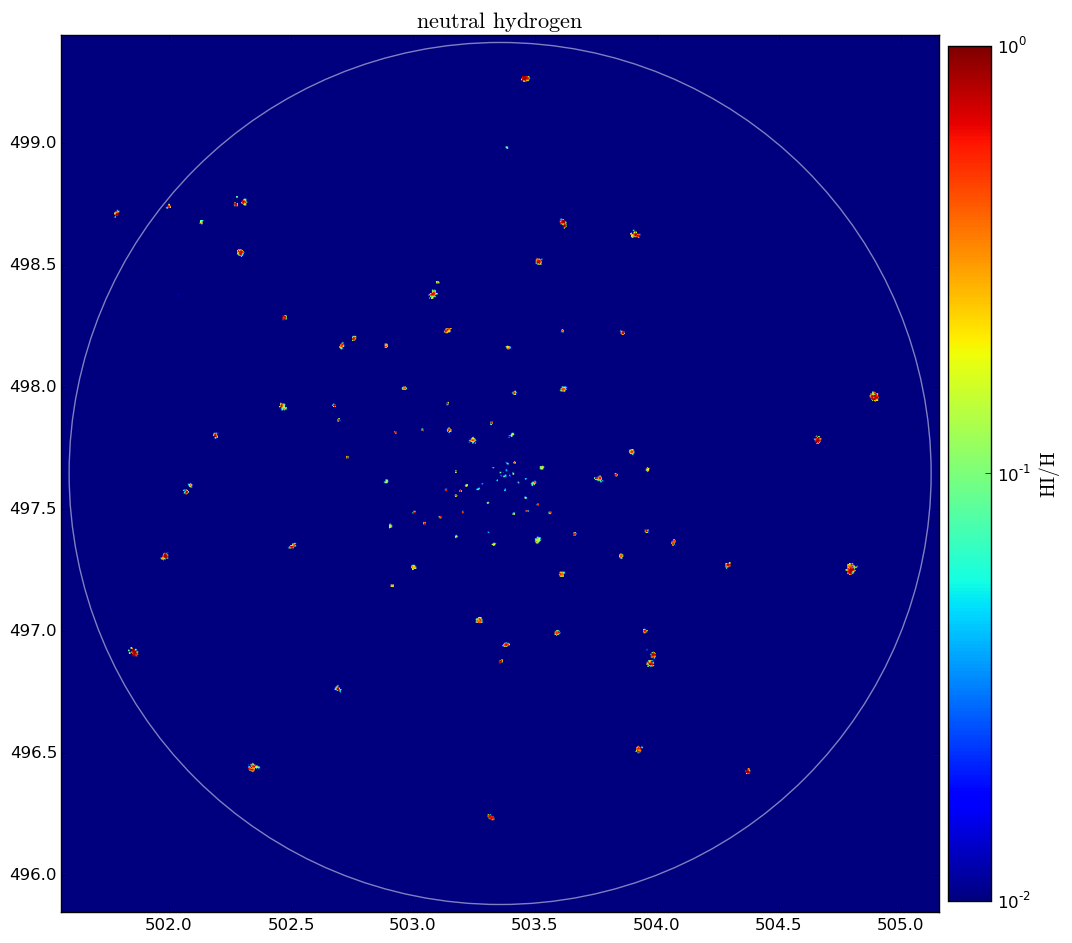}}
\caption{Distribution of neutral hydrogen in a very massive cluster of mass $M=1.28\times10^{15}~h^{-1}M_\odot$ at $z=0$. The neutral hydrogen fraction decreases quickly towards the center of the halo.}
\label{fig:Deficit_center}
\end{figure}

\begin{figure*}
\centerline{\includegraphics[width=0.8\textwidth]{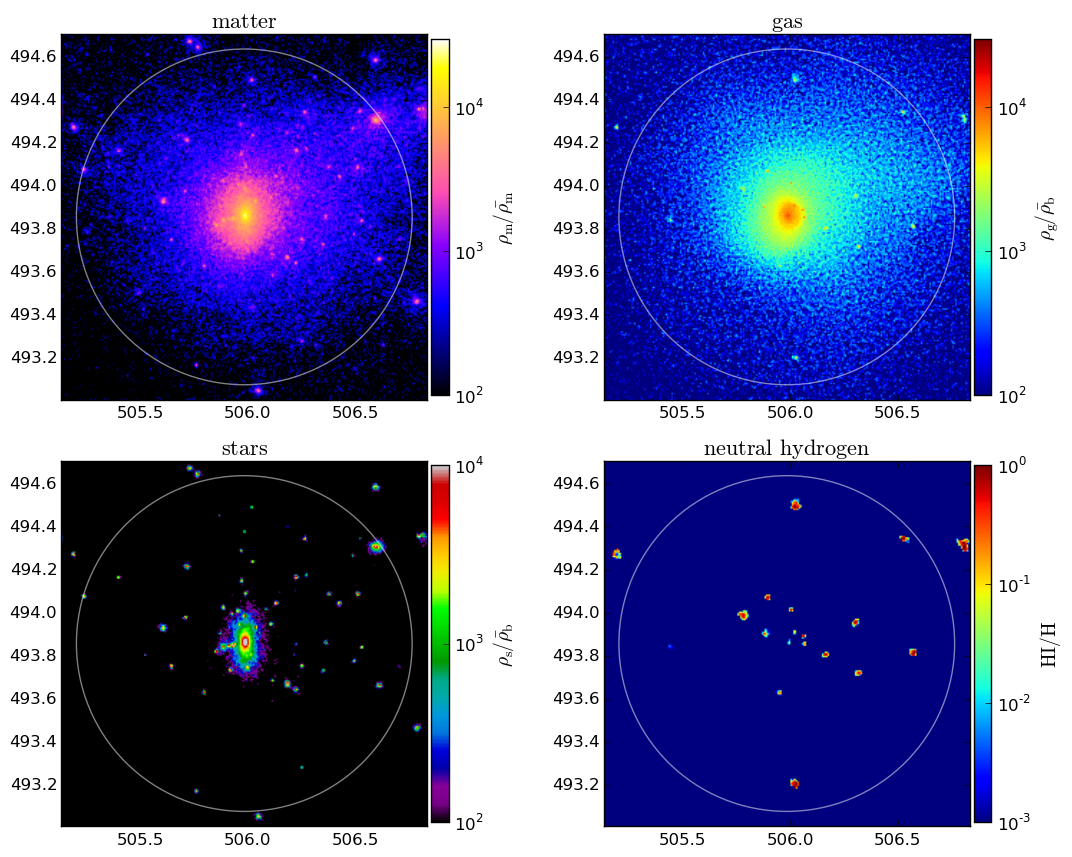}}
\caption{Effect of gas stripping in galaxy clusters. The panels show the spatial distribution of matter (top-left), gas (top-right), stars (bottom-left) and neutral hydrogen (bottom-right) centered on a halo of mass $M=1.1\times10^{14}~h^{-1}M_\odot$ at $z=0$. The x and y axis show the positions in $h^{-1}$Mpc; the width of the slices is equal to the diameter of the halo, which is $\sim1.5~h^{-1}$Mpc. It can be noticed that in the majority of the cases, the positions of the HI blobs are slightly displaced with respect to the positions of the galaxies. The white circles represent the halo $R_{200}$.}
\label{fig:Gas_stripping}
\end{figure*}

In order to further check our gas stripping hypothesis, we plot in Fig. \ref{fig:Gas_stripping} the spatial distribution of matter, gas, stars and neutral hydrogen for a halo of mass $M=1.1\times10^{14}~h^{-1}M_\odot$ from the AGN simulations at $z=0$. We find that the total number of subhalos identified by {\sc SUBFIND} is equal to 67. Among those, 52 host stars but only 2 have neutral hydrogen (one of them being the cD galaxy).  As can be seen, the positions of the neutral hydrogen blobs do not perfectly match, in the majority of the cases, the positions where stars are found. It is worth to point out that overdensities in the matter field have correspondence in the distribution of stars, in the same way as overdensities in the gas distribution match the positions of the HI blobs, but in general there is a small displacement among matter/stars and gas/HI. This just points out that whereas stars follow the same dynamics as CDM, gas behaves differently. 

In most of the cases, the positions of the HI blobs are just slightly displaced with respect to the positions where subhalos are located. This small displacement is however enough for {\sc SUBFIND} to fail at linking the gas to the subhalos, and indicates that gas stripping has taken place recently. We notice that since most of the galaxies do not have neutral hydrogen, {\sc SUBFIND} finds that the biggest subhalo of a given halo (that we refer to as the cD galaxy) hosts a very large amount of HI. The reason for this is that, since the spatial positions of the HI blobs and the galaxies are different, the neutral hydrogen blobs are accounted for in the largest subhalo, and not in the different galaxies. We find that in the $96\%$ and $91\%$ of the halos, the HI in their galaxies (mainly coming from the cD galaxy) account for more than $90\%$ and $95\%$ of the overall halo HI content, respectively. 

Unfortunately, the problems discussed above do not allow us to study the dependence of the HIMF on halo mass. As a very rough approximation, we plot in the bottom panel of Fig. \ref{fig:Number_galaxies} the average HI mass per galaxy, computed by dividing the overall HI mass in a given halo by the total number of subhalos within it, as a function of the halo mass. We warn the reader that the numbers quoted in the y-axis have to be taken with caution, since these results clearly depend on the resolution of the simulations: the higher the resolution the larger the number of subhalos it will be found, while the total HI mass is not expected to depend that strongly on resolution. In other words, the important thing in that plot is the trend rather than the absolute scale. Our results point out that the average HI mass in galaxies decreases with the mass of its host halo, although there is a rather large scatter in the trend. Therefore, HI poor galaxies are more likely to be found in clusters of galaxies than in groups. 

Even though in our simulations we can not directly link HI to galaxies, we can still compare our results against some observational results. For instance, it is known that the fraction of galaxies with HI detected by surveys declines towards the center of the halos \citep[see for instance][]{Hess_2013, Yoon_2015}. In Fig. \ref{fig:Deficit_center} we plot the spatial distribution of neutral hydrogen within a massive cluster of mass $M=1.28\times10^{15}~h^{-1}M_\odot$ at $z=0$ taken from the AGN simulations. It can be seen that towards the halo center the HI clouds are smaller and the average neutral hydrogen within them quickly decreases, the same trend seen in observations.

It is not obvious whether the gas blobs we find represent a physical situation \citep[e.g. gas stripping by the ICM; see for instance][]{2014ApJ...792...11J,2007ApJ...671..190S}, a numerical artifact of our hydrodynamic solver, where residual errors in our SPH scheme prevents the disruption of those HI blobs, or a mix of the two. In order to address this issue one could run simulations with different SPH methods or with grid methods, and investigate how results change. Besides, the stability of our result should be verified against resolution. This is however beyond the scope of this paper and we leave it for a future work.

Two important questions naturally then show up: 1) How important is the HI mass outside galaxies? and 2) How important is the HI mass arising from numerical artifacts on the overall HI content of halos?

\begin{figure*}
\centerline{\includegraphics[width=1.0\textwidth]{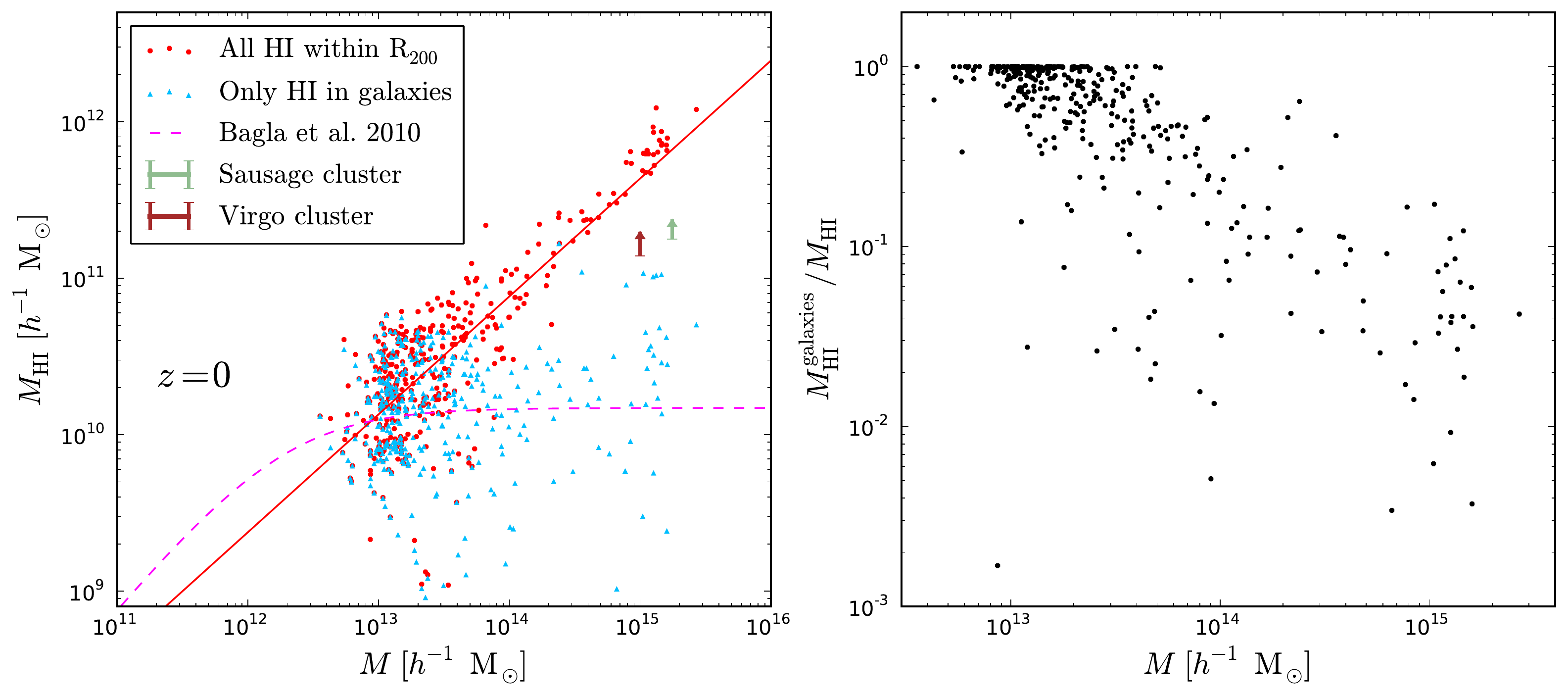}}
\caption{Neutral hydrogen mass outside galaxies. For each halo in the AGN simulations we have computed the HI mass within $R_{200}$ at $z=0$ . The blue points in the left panel show the results as a function of the halo mass together with the best-fit obtained by using the expression $M_{\rm HI}=e^\gamma M^\alpha$. For each halo we have also computed the HI mass residing only in galaxies, and the orange triangles in the left panel display the results. The green and brown arrows represent a lower limit on the neutral hydrogen mass hosted by the Sausage and Virgo clusters, respectively. The right panel shows the ratio, for each halo, between the HI mass in galaxies over the whole neutral hydrogen mass.}
\label{fig:HI_blobs}
\end{figure*}

In order to address the first question we have computed, for each halo of the AGN simulations at $z=0$, the HI mass within its $R_{200}$ which is located within galaxies. We have taken special care of the cD galaxy, since the output of {\sc SUBFIND} will assign to it most of the HI blobs. In order to circumvent this problem we have computed the HI mass of the cD galaxy as the HI mass contained in a sphere of radius of $50~h^{-1}$kpc\footnote{We notice that our results are robust against reasonable variations in the size of the cD galaxy.}. In the left panel of Fig. \ref{fig:HI_blobs} we show the results using this procedure together with the results taking all HI within $R_{200}$. We find that the HI mass hosted by galaxies in clusters is much lower than its overall HI mass, while for groups the difference is much smaller. The right panel of Fig. \ref{fig:HI_blobs} shows the ratio, for each halo, between the HI mass in galaxies to the overall HI mass. It can be seen that for groups, the HI mass in galaxies is very similar to the overall HI mass while for galaxy clusters the galaxies content in HI is a rather small fraction of the overall halo neutral hydrogen mass. In the left panel of Fig. \ref{fig:HI_blobs} we also display the prediction of the \citet{Bagla_2010} model for the $M_{\rm HI}(M,z)$ function. Our results point out that the \citet{Bagla_2010} model reproduces pretty well our measurements of the $M_{\rm HI}(M,z)$ function when the contribution of HI outside galaxies is removed, although the scatter is very large. We note however again that the Bagla model is inconsistent with the lower limit on the HI mass from the Sausage and Virgo clusters.

\begin{figure*}
\centerline{\includegraphics[width=1.0\textwidth]{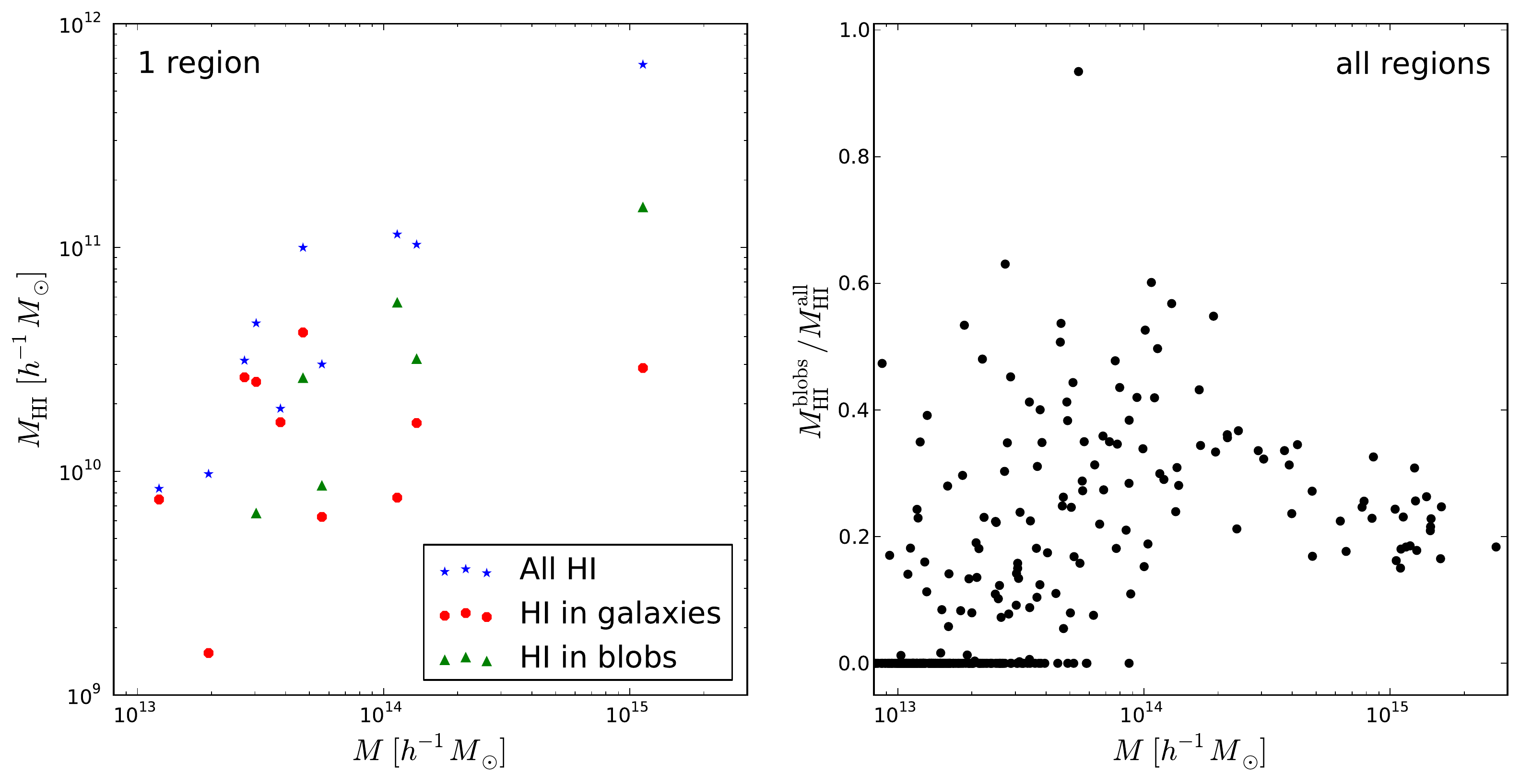}}
\caption{HI mass in numerical blobs. For each halo of the AGN simulations at $z=0$ we have computed all HI mass within $R_{200}$ (blue stars), the HI mass in galaxies (red points) and the HI in blobs (subhalos with no stars and dark matter; green triangles) and we show the results in the left panel. For clearness, we only display the results for one single region. The right panel shows the ratio between the HI mass in blobs to the overall HI mass for all halos.}
\label{fig:Kaufmann_blobs}
\end{figure*}

In order to address, at least partially, the question of how important is the HI mass in blobs whose nature is not physical but simply numerical, we have run {\sc SUBFIND}, on top of all of our resimulated regions at $z=0$ for the simulation with AGN feedback, allowing it to identify gas clouds as subhalos. We have then identified blobs of gas as subhalos with no CDM and stars particles in them. The reason for doing this is that we expect the numerical blobs to show up as isolated and concentrated clouds of gas. Therefore {\sc SUBFIND} is likely to identify these structures as purely gaseous subhalos, while stripped gas will be more diffuse and thus {\sc SUBFIND} will not recognize it as subhalos. In Fig. \ref{fig:Kaufmann_blobs} we show, for each halo, the overall HI mass within $R_{200}$ together with the HI mass within galaxies and within the blobs. 

We find that for groups, the HI mass in numerical blobs is just a small fraction of the overall HI mass, while for galaxy clusters the HI mass in those blobs is much larger and surpasses the HI mass within galaxies. In the right panel of Fig. \ref{fig:Kaufmann_blobs} we plot, for each halo, the ratio between the HI mass in numerical blobs over the overall HI mass. We find that the contribution to the overall HI mass from numerical blobs (notice that we are assuming that all purely gaseous subhalos are numerical, which may not be true) is below the $50\%$ in the majority of the cases. Thus, we conclude that the results of this paper are robust against numerical problems and the HI masses we quote can be trusted within a factor of 2 for galaxy clusters while the HI mass within groups is much less sensitive to these issues. 

We however emphasize that a more detailed study is required in order to more closely address the nature of the isolated HI blobs we find in our simulations. Notice that HI clouds without optical counterparts have been found in \citet{Davies_2004,Minchin_2005, Kent_2007, Koopmann_2008, Rhys_2012a,Rhys_2012b, Janowiecki_2015}. We plan to address this in a future paper.

\end{appendix}

\bibliographystyle{mnbst}
\bibliography{references}

\end{document}